\documentclass[final,journal]{IEEEtran}

\usepackage{amssymb}
\usepackage{amsmath,bm,amsthm}
\usepackage{graphicx}
\usepackage{psfrag}
\usepackage{epsf,epsfig}
\usepackage{epstopdf}
\usepackage{subfigure}
\usepackage{cite,color}
\usepackage{multirow}

\definecolor{red}{rgb}{1,0,0}  

\newcommand{\rmq}{{\rm q}}

\newcommand{\Tr}{{\rm Tr}}
\newcommand{\expect}[1]{{{\mathbb E}\left[#1\right]}}   
\newcommand{\var}[1]{{{\rm Var}\left[#1\right]}}   
\newcommand{\etr}{{\mathrm{etr}}}

\newcommand{\Scal}{\mathcal{S}} 
 
\newcommand{\Ncal}{\mathcal{N}}
  
\newcommand{\Ccal}{\mathcal{C}} 
\newcommand{\Ucal}{\mathcal{U}} 
\newcommand{\Vcal}{\mathcal{V}} 
\newcommand{\Gcal}{\mathcal{G}} 
\newcommand{\Mcal}{\mathcal{M}} 
\newcommand{\Tcal}{\mathcal{T}} 
\newcommand{\Hcal}{\mathcal{H}}

\newcommand{\abf}{\bm{a}}
\newcommand{\bbf}{\bm{b}}
\newcommand{\Abf}{\bm{A}}
\newcommand{\Bbf}{\bm{B}}

\newcommand{\Ebf}{\bm{E}}
\newcommand{\ebf}{\bm{e}}
\newcommand{\Ybf}{\bm{Y}}
\newcommand{\Cbf}{\bm{C}}
\newcommand{\Xbf}{\bm{X}}
\newcommand{\Ibf}{\bm{I}}

\newcommand{\Zbf}{\bm{Z}}
\newcommand{\ybf}{\bm{y}}
\newcommand{\Hbf}{\bm{H}}

\newcommand{\xbf}{\bm{x}}
\newcommand{\Ubf}{\bm{U}}
\newcommand{\Vbf}{\bm{V}}
\newcommand{\Mbf}{\bm{M}}
\newcommand{\Nbf}{\bm{N}}
\newcommand{\Rbf}{\bm{R}}
\newcommand{\Pbf}{\bm{P}}
\newcommand{\cbf}{\bm{c}}

\newcommand{\Lbf}{\bm{L}}
\newcommand{\Qbf}{\bm{Q}}
\newcommand{\Sbf}{\bm{S}}
\newcommand{\Tbf}{\bm{T}}
\newcommand{\Omegabf}{\bm{\Omega}}
\newcommand{\Deltabf}{\bm{\Delta}}
\newcommand{\thetabf}{\bm{\theta}}
\newcommand{\phibf}{\bm{\phi}}
\newcommand{\Thetabf}{\bm{\Theta}}

\newcommand{\Cbb}{\mathbb{C}}

\newcommand{\dM}{{\rm dim}} 
\newcommand{\erf }{{\rm erf }}
\newcommand{\diag }{{\rm diag }}

\usepackage{stackengine}
\newcommand\barbelow[1]{\stackunder[1.2pt]{$#1$}{\rule{.8ex}{.075ex}}} 

\newcommand{\rhol}{\barbelow{\varrho} }  
 \newcommand{\rhou}{\bar{\varrho}}

\newcommand{\grass}[3]{ {\Gcal}_{#2,#3}^{ \mathbb #1 }} 
\newcommand{\stief}[3]{ {\Vcal}_{#2,#3}^{ \mathbb #1 }}

\newtheorem{Thm}{Theorem}
\newtheorem{Cor}{Corollary}
\newtheorem{Def}{Definition}
\newtheorem{Prop}{Proposition}
\newtheorem{Lem}{Lemma}
\newtheorem{Rem}{Remark}
\newtheorem{Example}{Example}

\begin{document}

\title{Density of Spherically-Embedded Stiefel and Grassmann Codes}

\author{
\IEEEauthorblockN{Renaud-Alexandre Pitaval, Lu Wei, Olav Tirkkonen, and Camilla Hollanti}

\thanks{
\copyright 2017 IEEE. Personal use of this material is permitted. Permission from IEEE must be obtained for all other uses, in any current or future media, including reprinting/republishing this material for advertising or promotional purposes, creating new collective works, for resale or redistribution to servers or lists, or reuse of any copyrighted component of this work in other works.
 
This work was supported in part by the Academy of Finland (grants \#276031  and \#299916) and by the Nokia Foundation. 
Part of this work was presented at the 2011 IEEE International Symposium on Information Theory and the 2012 Asilomar Conference of the Signals, Systems and Computers.

R.-A. Pitaval was with the Department of Communications and Networking, Aalto University, FI-00076 Espoo, Finland. He is now with Huawei Technologies Sweden AB, SE-164 94 Kista, Sweden (e-mail: renaud-alexandre.pitaval@alumni.aalto.fi).

L. Wei is with the Department of Electrical and Computer Engineering, University of Michigan-Dearborn, Michigan 48128, USA (e-mail: luwe@umich.edu). 

O. Tirkkonen is  with the Department of Communications and Networking, Aalto University, FI-00076 Espoo, Finland (email: olav.tirkkonen@aalto.fi).

C. Hollanti is with the Department of Mathematics and Systems Analysis, P.O. BOX 11100, FI-00076 AALTO, Finland (email: camilla.hollanti@aalto.fi).

}
}

\maketitle
\vspace{-1cm}
\begin{abstract} 
The density of a code is the fraction of the coding space covered by
packing balls centered around the codewords. A high density  
indicates that a code performs well when used as a uniform
point-wise discretization of an ambient space. This paper investigates
the density of codes in the complex Stiefel and Grassmann manifolds
equipped with 
the chordal distance arising from an Euclidean embedding, 
including the unitary group as a special case. The choice of distance 
enables the treatment of the manifolds as subspaces of Euclidean
hyperspheres. In this geometry, the densest packings are not
necessarily equivalent to   
maximum-minimum-distance codes. Computing a code's density follows
from computing: i) the normalized volume of a metric ball and ii) the
kissing radius, the radius of the largest balls one can pack around
the codewords without overlapping. 
First, the normalized volume of a metric ball is evaluated by
asymptotic approximations. 
The volume of a small ball can be well-approximated by the volume of a
locally-equivalent tangential ball. In order to properly normalize
this approximation, the precise volumes of the manifolds induced by
their spherical embedding are computed. For larger balls, a
hyperspherical cap approximation is used, which is justified by a
volume comparison theorem showing that the normalized volume of a ball
in the Stiefel or Grassmann manifold is asymptotically equal to the
normalized volume of a ball in its embedding sphere as the dimension
grows to infinity. Then, bounds on the kissing radius are derived
alongside corresponding bounds on the density. Unlike  
spherical codes or 
codes in flat spaces, the kissing radius of 
Grassmann or Stiefel codes cannot be exactly determined from its
minimum distance. It is nonetheless possible to derive bounds on
density as functions of the minimum distance. Stiefel and Grassmann
codes have larger density than their image spherical codes when
dimensions tend to infinity. Finally, the bounds on density lead to
refinements of the standard Hamming bounds for Stiefel and Grassmann
codes.
\end{abstract}

\IEEEpeerreviewmaketitle

\section{Introduction}

Sphere packing is a classical problem with a long history from
geometry to information
theory~\cite{Rankin,Rogers,toth1965,SloaneBook}. In his 1948 seminal
work~\cite{Shannon1948}, Shannon made the connection between the
capacity of an additive white Gaussian noise channel and packing of
multidimensional spheres. This interpretation was later generalized to
non-coherent multi-antenna  
channels and packings in products of Grassmann manifolds~\cite{Tse}.

Stiefel and Grassmann codes are matrix codes with applications to
Multiple Input Multiple Output (MIMO)
communications~\cite{Agrawal,Tse,overview,TWC14}, Code Division
Multiple Access (CDMA) wireless systems~\cite{HeathCDMATIT06}, and
compressive
sensing~\cite{TroppTIT04,calderbank2015block}. The
complex Stiefel manifold is the space of rectangular semi-unitary
matrices. 
The Grassmann manifold is the space of eigenspaces spanned by the Stiefel matrices. An element in the Grassmann manifold is an equivalence class of Stiefel matrices, which can be alternatively represented by a unique projection matrix. When discussing codes in Grassmann manifolds of one-dimensional subspaces, the equivalent language of frame theory is often used~\cite{calderbank2015block}.

Depending of the application and convenience, several non-equivalent distances can be defined on these spaces~\cite{conway-1996-5,Edelman}. A common distance for the  unitary group and Stiefel manifold is simply the Frobenius norm of the difference of two matrices~\cite{Bachoc08}. Similarly for the Grassman manifold, an often-used distance arises from the Frobenius norm of the difference between two projection matrices~\cite{overview,AshikhminCalderbankTIT10}.
These  distances correspond to embed the manifolds into Euclidean spaces. 
Through these embeddings,  each manifold is a subset of an hypersphere and the distance is the length of a chord. 

For the application of limited-feedback MIMO precoding, unitary precoder designs have been reduced to quantization problems on Grassmann and Stiefel manifolds with the associated chordal distance~\cite{overview,Dai08,TWC14}. In the context of space-time coding, the code performance in the low signal-to-noise-ratio (SNR) regime has also been related to the chordal distance~\cite{HanRosenthalGeo,AshikhminCalderbankTIT10}. More recently, analogical results have been derived for linear classification:  In the small model mismatch regime,  the number of classes that can be distinguished by a subspace classifier is governed by the chordal distance among subspaces, and more generally by their principal angles~\cite{HuangQiuCalderbankTSP16}.   
 
In the last decade, basic coding-theoretic results estimating the relationship between the cardinality and the minimum distance of codes in Grassmann and Stiefel manifolds have been published~\cite{Barg,Barg2,henkel,Han,Bachoc06,Bachoc08,CreignouISIT,Dai08,Krishnamachari_R,Krishnamachari_C}. The standard Hamming bound relates the minimum distance to the notion of code density. 
The density of a code is the maximum portion of the coding space covered by non-intersecting balls of equal radius. 
In classical geometry, maximizing the code's density is known to be equivalent to maximizing its minimum distance. For the Stiefel and Grassmann codes measured by chordal distance, this equivalence does not always hold. 

The present paper investigates this surprising fact and discusses the density of codes in Grassmann and Stiefel manifolds equipped with their chordal distance. We provide  a direct connection between minimum distance and density via lower and upper bounds on the density for a given distance. This connection leads to new conjectured bounds on the minimum distance as a function of code cardinality. Related to applications, numerical evaluations suggest that  the density may be a more relevant criterion for MIMO precoding than the minimum distance. While for space-time coding, the minimum distance of a code may be more important than its density.

There are two main difficulties in evaluating the density of codes in these spaces: 1) evaluating the normalized volume of a ball, and 2) estimating the kissing radius of codes. 

The problem of estimating the volume of a metric ball has been addressed in~\cite{MukkRank1,Barg,henkel,Han,Dai08,DaiGrowth07,Krishnamachari_R,Krishnamachari_C}. In the large code cardinality regime, balls are small and can be approximated to be balls in flat space. While some exact evaluations were obtained in Grassmannian cases~\cite{MukkRank1,Dai08}, the case of the Stiefel manifold has been less addressed. 
A  powerful and general framework is provided in~\cite{henkel, Krishnamachari_R,Krishnamachari_C}. 
However, it appears that the state-of-the-art volumes in the literature do not correspond to the desired metrics. Indeed, the volume element is unique up to a non-vanishing scaling factor which  is often dismissed, as  it can be absorbed in the overall normalization. From Nash embedding theorem~\cite{Nash}, every Riemannian metric can be seen as induced by an appropriate Euclidean embedding. There exists an intrinsic metric locally equivalent to the chordal distance, and thus a consistent volume density that defines the  notion of  volume on the manifold. 
In this respect,  we provide the exact scaling of the volume for Grassmann and Stiefel manifolds induced
by the spherical embedding, leading in turn to precise small ball approximations. 

For larger radius, we show that  the volume of  a ball in the manifold can be well approximated by  the ``area'' of the hyperspherical cap the ball is embedded  into. The approximation is supported by a volume comparison theorem showing that the normalized volume of a ball in the manifold is asymptotically equal to the normalized volume of a ball in the embedding sphere.
This result generalizes and provides  a structural unification of our previous results in~\cite{WPTC_ISIT15,WPCT15,PWTC15}. 
The derivation is a by-product of the asymptotic Gaussianity of the chordal distance arising through its reduction to linear statistics. 
The intuition behind relates to a classical result by Borel~\cite{Borel} who proved that coordinates of a hypersphere are asymptotically Gaussian as the dimension tends to infinity~\cite{Richards}, and the long history of central limit theorems showing the Gaussian behavior of linear statistics~\cite{breuer2013central} in random matrix theory. 
 
Next, the paper addresses the evaluation of the kissing radius and density of Grassmann and Stiefel codes. The kissing radius is the analog of the packing radius for  linear codes~\cite{SolePAckingRadius}, which has applications in, e.g., sphere-decoder optimization~\cite{SchenkLetter09,SchenkSCC10}. It is the largest possible radius of packing balls around the codewords of a code. 
The kissing radius also relates to rate-distortion theory as it is the smallest possible  distance from a codeword to the border of its Voronoi cell.  
The problem reduces to finding the minimum mid-distance between two points at a given distance $\delta$. For a geodesic distance  on a flat space, the answer is simply $\delta/2$, the so-called packing radius. With a strictly extrinsic distance, the triangle inequality is never satisfied with equality and the kissing radius is greater than $\delta/2$. While for spherical codes there is a one-to-one mapping between the kissing radius and the minimum distance, this is not always the case for the Grassmann and Stiefel codes with chordal distance. 
As a consequence, the density is not a single-variable function of the minimum distance of the code, and two codes with equal minimum distance could have different densities. This is in sharp contrast with classical packing problems where maximizing the density is equivalent to maximizing the minimum distance. 
The kissing radius and the density cannot be determined solely from the minimum distance but it is possible to derive bounds. Combining these bounds with the volume comparison theorem discussed above shows that the densities of Grassmann and Stiefel codes are asymptotically greater than or equal to the densities of their image spherical codes.   The bounds are shown to be tight by simulations.


Finally, a direct application of bounds on density is to revisit the Hamming bounds. 
The results  of this paper improve the Hamming bound for the Grassmann case in~\cite{Barg,henkel}; and generalize the bound  for the unitary group in~\cite{Han} to the Stiefel manifold. 
	
The rest of this paper is organized as follows. Section II introduces the considered spaces and their geometry. Section III states the problem and necessary definitions. 
Section IV addresses  the problem of the volume of a ball. In Section V, bounds on kissing radius and density are derived.   Section VI provides concluding remarks. 

\section{The Grassmann and Stiefel Manifolds}
We consider the following Riemann manifolds equipped with a chordal distance induced by their canonical spherical embedding. 
Throughout the paper, $\Mcal$ stands for the unitary group $\Ucal_n$, Stiefel manifold $\stief{C}{n}{p}$, or Grassmann manifold $\grass{C}{n}{p}$ embedded into the sphere $\Scal^{D-1}(R)$ of radius $R= \sqrt{n} $,  $\sqrt{p}$ or ${ \scriptscriptstyle \sqrt{ \frac{p(n-p)}{2n}}}$, in a Euclidean space of dimension $D=2n^2$, $2np$ or $n^2-1$,  respectively. 
Elements in the manifold are represented by matrices in $\mathbb{C}^{n
  \times n}$ or $\mathbb{C}^{n \times p}$ endowed by the inner product
$\langle \cdot,\cdot \rangle = \mathfrak{R} \Tr \cdot^H \cdot$, where
$\mathfrak{R}$ is the real part and $\Tr$ is the matrix trace. The
geometric description below follows directly by
generalizing~\cite{conway-1996-5} and~\cite{Edelman} to the complex
space. 
The chordal distances,  
the dimension $\dM$ of the manifolds, and the
corresponding embeddings  
are summarized in Table~\ref{tab:emb}.

\begin{table}[t]
\caption{Manifolds of dimension $\dM$ with their spherical embeddings: $(\Mcal,d_c) \hookrightarrow  \Scal^{D-1}(R)$. \label{tab:emb}}
\vspace{-0.5cm}
\begin{center}
\begin{tabular}{c|c|c|c|c}	
$\Mcal$ & $\dM$& $d_c(\Ubf,\Vbf)$& $D$ & $R$ \\\hline
$\Ucal_n$ & $n^2$& $\|\Ubf-\Vbf \|_F$  & $2n^2$ &  $ \sqrt{n}$ \\
$\stief{C}{n}{p}$ & $2np-p^2$&$\|\Ubf-\Vbf \|_F$  &  $2np$& $ \sqrt{p}$ \\
$\grass{C}{n}{p}$ &$2np-2p^2$& $\frac{1}{\sqrt{2}} \|\Ubf\Ubf^H-\Vbf\Vbf^H \|_F$  &  $n^2-1$ & ${  \sqrt{ \frac{p(n-p)}{2n}}}$ \\
\end{tabular}
\end{center}
\end{table}

\subsection{Hypersphere}
The Euclidean $(D-1)$-\emph{sphere} of radius $R$ in $\mathbb{R}^D$ is defined as
\begin{equation} \Scal^{D-1}(R)= \left\{ \xbf \in \mathbb{R}^D \quad | \quad ||\xbf||_2= R \right\}.\end{equation} 
For $R=1$, 
we simply write $\Scal^{D-1}$.  
The \emph{chordal distance} is the natural Euclidean distance applied to elements on the sphere. 
Given $\xbf,\ybf \in\Scal^{D-1}(r) $ it is simply the norm of the difference:
\begin{equation}
d_c(\xbf,\ybf) = ||\xbf - \ybf ||_2.
\end{equation}
It is an extrinsic distance as it measures the length of a chord outside of the manifold itself, which here is the surface of the sphere. 

\subsection{Unitary Group}
The \emph{unitary group} is the set of unitary matrices, 
\begin{equation} \Ucal_n  = \left\{\Ubf \in \mathbb{C}^{n\times n} \quad | \quad \Ubf^H \Ubf = \Ubf \Ubf^H = \Ibf_n \right\},
\end{equation}
where $ \cdot^H$ denotes the Hermitian conjugate. This compact Lie group is a manifold of dimension $\dim \Ucal_n = n^2$.  
By differentiating the unitary constraint, one can verify   
that the tangent space $\Tcal_{\Ubf} \Ucal_n$  at $\Ubf$  is the set of matrices $\Deltabf \in \mathbb{C}^{n\times n} $  such that $\Ubf^H \Deltabf$ is skew-Hermitian. 
Specifically, at identity, the tangent space  $\Tcal_{\Ibf} \Ucal_n $ is the Lie algebra of skew-Hermitian $n \times n$ matrices  $ \mathfrak{u}(n)$. 

One can parametrize $\Ucal_n$ with reference to a fixed $\Ubf\in\Ucal_n$ via 
skew-Hermitian matrices as $\Vbf= \Ubf \exp(\Ubf^H\Deltabf) \in
\Ucal_n $ with $\Ubf^H \Deltabf \in \mathfrak{u}(n) $. For a fixed
$\Deltabf$, the exponential map $\exp(\cdot)$ defines the geodesic
between $\Ubf$ and $\Vbf$ by mapping the tangent space to the manifold
as $\Vbf(t)= \Ubf \exp(t\ \Ubf^H\Deltabf )$ with $0\leq t \leq 1$. A
Riemannian metric may be defined from the canonical embedding of
$\Ucal_n$ in the Euclidean space $(\mathbb{C}^{n \times n}, \langle
\cdot, \cdot \rangle)$, then $\exp(\cdot)$ is the matrix exponential.

Consider the eigenvalue decomposition 
$\Ubf^H \Vbf = \Omegabf \diag(e^{i \theta_1},e^{i \theta_2},\ldots,e^{i \theta_n})\Omegabf^H$  
where the diagonal elements of $\Omegabf \in \Ucal_n$  are
non-negative and real.  
This decomposition is unique if  
these \emph{principal angles} can be strictly ordered $\pi \geq \theta_n >  \ldots >  \theta_2 > \theta_1\geq -\pi$. This  leads to the corresponding eigenvalue decomposition $\Ubf^H\Deltabf =  \Omegabf \diag(\theta_1,\theta_2,\ldots,\theta_n)\Omegabf^H $, and accordingly the geodesic between $\Ubf$ and $\Vbf$ becomes
\begin{equation} 
\label{eq:UnEigDecomp}
\Vbf(t) = \Ubf \Omegabf \diag(e^{i \theta_1 t},e^{i \theta_2t},\ldots,e^{i \theta_nt})\Omegabf^H.
\end{equation}

The geodesic distance is the intrinsic distance between two points obtained by integrating the geodesic path along the manifold.  It is given by the norm of the  tangent direction  according to the considered Riemannian metric 
\begin{equation}  
d_g(\Ubf, \Vbf) = \| \Deltabf \|_F=\|\Ubf^H \Deltabf \|_F = \sqrt{\sum_{i=1}^n \theta_i^2 }.
\end{equation}

Alternatively, the canonical distance in the ambient matrix-space is 
\begin{eqnarray}
d_c(\Ubf, \Vbf) &=& \| \Ubf-\Vbf \|_F = \sqrt{2n - 2\mathfrak{R}\Tr(\Ubf^H\Vbf) } \nonumber\\
&=& \sqrt{ 4 \sum_{i=1}^n \sin^2 \frac{\theta_i}{2} }  
\end{eqnarray} 
where the last equality follows from the decomposition~\eqref{eq:UnEigDecomp}. 

By observing that  $d_c(\mathbf{0}, \Vbf)=\| \Vbf \|_F = \sqrt{n} $ for all $\Vbf \in\Ucal_n $, one verifies that the unitary group equipped with $d_c$ has an isometric embedding into the sphere $\Scal^{2n^2-1}(\sqrt{n})$. The concrete embedding $(\Ucal_n,d_c) \hookrightarrow \Scal^{2n^2-1}(\sqrt{n})$ is obtained from the classical mapping $\mathbb{C}^{n \times n} \hookrightarrow  \mathbb{R}^{2n^2}$  by  vectorizing a complex matrix into a real vector. 

\subsection{Stiefel Manifold}

The complex \emph{Stiefel manifold} is the space of rectangular semi-unitary matrices,
\begin{equation}  \stief{C}{n}{p}  = \left\{\Ybf \in \mathbb{C}^{n\times p} \quad | \quad \Ybf^H \Ybf = \Ibf_p \right\} .
\label{eq:StiefDef}
\end{equation}
This provides a generalization of both the hypersphere and the unitary group. For $p=n$, one directly recovers $\stief{C}{n}{n}= \Ucal_n$, while for $p=1$ and by identification of $\mathbb{C}^{n}$ with $\mathbb{R}^{2n}$, this corresponds to the unit sphere $\stief{C}{n}{1} = \Scal^{2n-1}$.

As for the unitary group, the tangent space at $\Ybf$ is the set $\Tcal_{\Ybf} \stief{C}{n}{p}$ of matrices $\Deltabf \in \mathbb{C}^{n \times p}$ such that $\Ybf^H \Deltabf \in \mathfrak{u}(p) $ is skew-Hermitian. 
In general, tangents have the form $\Deltabf = \Ybf \Abf + \Ybf_{\bot} \Bbf$  where $\Abf$ is $p \times p $ skew-Hermitian,  $\Ybf_{\bot}$ is the orthogonal complement of $\Ybf$ such that the concatenation $(\Ybf \, \Ybf_{\bot}) \in \Ucal_n$ is unitary, and $\Bbf$ is an $(n-p) \times p $ arbitrary complex matrix. 
Specifically at ``identity'' $\Ibf_{n,p} \triangleq \begin{pmatrix} \Ibf_p \\ \mathbf{0}\end{pmatrix}$, tangents are of the form 
\begin{equation} \Deltabf= \begin{pmatrix} \Abf \\ \Bbf \end{pmatrix} \in  \mathbb{C}^{n \times p} \text{  with  }  \Abf \in \mathfrak{u}(p) . 
\end{equation}
By counting the degrees of freedom in the tangent space, one finds that the Stiefel manifold is a space with dimension  $\dim \stief{C}{n}{p}  = 2np-p^2$. 

Given a starting point $ \Ybf = \Ybf(0)$ and a fixed tangent direction $\Deltabf= \Ybf \Abf + \Ybf_{\bot} \Bbf \in \Tcal_{\Ybf} \stief{C}{n}{p}$,  the canonical embedding of $\stief{C}{n}{p}$ in the Euclidean space $(\mathbb{C}^{n \times p},\langle \cdot, \cdot\rangle )$  leads  to the following geodesic equation~\cite{Edelman}, 
\begin{equation} \Ybf(t) = \begin{pmatrix} \Ybf &\Deltabf\end{pmatrix} 
\label{eq:StiefGeo}
\exp t  \begin{pmatrix}\Ybf^H \Deltabf & - \Deltabf^H  \Deltabf \\ \Ibf_p & \Ybf^H \Deltabf \end{pmatrix}  \Ibf_{2p,p} e^{-t \Ybf^H \Deltabf}. 
\end{equation} 
The  geodesic distance 
between $\Ybf$ and $ \Zbf= \Ybf(1)$ is then
\begin{equation}  
d_g(\Ybf, \Zbf) = \| \Deltabf \|_F =  \left( \| \Abf \|_F^2 + \| \Bbf \|_F^2 \right)^{1/2}. 
\end{equation}  
\begin{Rem} \label{Rem:RiemLogMap}
To the best of our knowledge, when $p \neq n$,  unlike for the unitary group, it is  not known how to compute the geodesic $\Ybf(t)$  in closed-form, given only $ \Ybf =\Ybf(0) $ and $\Zbf =\Ybf(1) $.
\end{Rem}
The corresponding Euclidean/chordal distance from the ambient space is
\begin{equation} 
d_c(\Ybf, \Zbf) = \| \Ybf-\Zbf \|_F  .
\end{equation}  
Similarly as for the unitary group, one can verify that $d_c(\mathbf{0}, \Zbf)=\sqrt{p}$ for all $\Zbf$, and thus this gives an  isometric embedding of the Stiefel manifold $\stief{C}{n}{p}$  into $ \Scal^{2np-1}(\sqrt{p})$. 

As an alternative to~\eqref{eq:StiefDef}, the Stiefel manifold can be
treated as the quotient space $\stief{C}{n}{p}  \cong \Ucal_n /
\Ucal_{n-p}$, where a point in  $\stief{C}{n}{p}$ is an  
equivalence class of unitary matrices \mbox{$[\Ubf] = \left \{\Ubf  \begin{pmatrix} {\bm I}_p & {\bm 0} \\ {\bm 0} & \Omegabf  \end{pmatrix}   \; \bigg| \; \Omegabf \in \Ucal_{n-p}  \right\} $}. A natural geometry in this interpretation of the  Stiefel manifold is the one inherited from the geometry of the unitary group $\Ucal_n$ embedded in $(\mathbb{C}^{n \times n}, \langle \cdot, \cdot\rangle)$.
One can split the tangent space  of the unitary group at $\Ubf$ between the so-called vertical and horizontal spaces: $\Tcal_{\Ubf} \Ucal_n = \Hcal_{\Ubf} \oplus \Vcal_{\Ubf} $. 
The vertical space $ \Vcal_{\Ubf} $ is the tangent space of $[\Ubf]\hookrightarrow \Ucal_n$ at $\Ubf$  corresponding to ``movements'' inside the equivalence class. 
The horizontal space  $ \Hcal_{\Ubf}$ is the orthogonal complement of the vertical space, providing a unique representation for tangents to the quotient space. For  $ \stief{C}{n}{p} \cong  \Ucal_n / \Ucal_{n-p}$ at $\Ubf \in \Ucal_n  $, it is the collection of matrices  $\Deltabf_{*} =  \Ubf \begin{pmatrix}\Abf  & -\Bbf^H \\ \Bbf & \mathbf{0} \end{pmatrix}$, where $\Abf\in \mathfrak{u}(n)$ and $\Bbf \in \mathbb{C}^{(n-p) \times p}$.
With this non-equivalent geometry, the geodesic distance between $ \Ubf$ and $ \Vbf=\Ubf\exp(\Ubf^H\Deltabf_*)$ is
$
d_{g*}(\Ubf, \Vbf) = \| \Deltabf_* \|_F =   \left( \| \Abf \|_F^2 + 2\| \Bbf \|_F^2 \right)^{1/2}.
$ The metric induced by this embedding,  
and the resulting volume is discussed in~\cite{Krishnamachari_C}. This
metric is  not further considered in this paper.

\subsection{Grassmann Manifold}
The complex \emph{Grassmann manifold} $\grass{C}{n}{p}$ is the quotient space of $\stief{C}{n}{p}$ over $\Ucal_p$:
\begin{equation}  \grass{C}{n}{p}  \cong \stief{C}{n}{p}/\Ucal_p \subset \mathbb{C}^{n \times p} .
\end{equation}  
Elements in $\grass{C}{n}{p}$ are  equivalence classes of rectangular semi-unitary matrices $\Ybf \in \stief{C}{n}{p}$:
\begin{equation} [\Ybf] = \{\Ybf \Qbf \;|\; \Qbf \in \Ucal_p \} .
\end{equation} 
We can identify the equivalence class with a unique matrix representation, which is desirable for pratical implementation and computation.  
By symmetry (c.f. the discussion in~\cite{PWTC15}) and for simplicity, we will assume  $p\leq n/2$  for the Grassmann manifold, all along this paper. 

As described above, tangent vectors at $\Ybf \in \stief{C}{n}{p} $ take the form \mbox{$\{ \Ybf \Abf + \Ybf_{\bot} \Bbf \}  $}. This can be again split between  the vertical space $\Vcal_{\Ybf} =\{ \Ybf \Abf  \;|\;  \Abf\in \mathfrak{u}_p \}$ and the horizontal space  $\Hcal_{\Ybf}  =\{   \Ybf_{\bot} \Bbf \;|\;   \Bbf \in  \mathbb{C}^{n\times(n-p)} \} $. 
At $\Ibf_{n,p}$, horizontal tangents are thus of the form $\Deltabf = \begin{pmatrix} {\bm 0} \\ \Bbf \end{pmatrix}$, and  we have $\dim  \grass{C}{n}{p}  = 2p(n-p)$. 
Consider the compact singular value decomposition  $\Bbf = \Lbf \Thetabf \Rbf^H $, where $\Lbf \in \stief{C}{n-p}{p} $,  $\Rbf \in \Ucal_{p} $ and  $\Thetabf = \diag(\theta_1,\ldots,\theta_p)$.
By setting $\Abf = \bm{0}$, the geodesic equation~\eqref{eq:StiefGeo} for the Grassmann manifold reduces to~\cite{Edelman}
\begin{equation}
\label{eq:GrassGeodesicEq}
 \Ybf(t) =   \Ybf\Rbf \cos ( \Thetabf t ) \Rbf^H  + \Ybf_\bot \Lbf \sin (\Thetabf t) \Rbf^H .
\end{equation}  
The geodesic equation could be rotated from the right by any unitary
matrix as it will still belong to the same equivalence class. 
Without loss of generality 
one can thus restrict  
the range of the singular
values of $\Bbf$ to $0\leq \theta_i \leq \frac{\pi}{2}$ for all $i$.
These values are known as the  
\emph{principal angles} between the planes $[\Ybf(0)]$ and
$[\Ybf(1)]$. Contrary to the Stiefel
manifold, here given two end points $\Ybf=\Ybf(0)$ and $\Zbf$, one can
compute the tangent in the geodesic~\eqref{eq:GrassGeodesicEq} and
satisfy $[\Zbf]=[\Ybf(1)]$ by singular value decomposition of $\Zbf^H
\Ybf $ and $\Zbf^H \Ybf_\bot$, where the singular values of $\Zbf^H
\Ybf $ are $\cos \theta_1,\ldots,\cos \theta_p$.

The geodesic distance is given by 
\begin{equation} 
d_g([\Ybf], [\Zbf]) =\| \Deltabf \|_F  =\|\Bbf\|_F =   \sqrt{ \sum_{i=1}^p \theta_i^2 }.
\label{eq:dgG}
\end{equation}
In~\cite{conway-1996-5}, the following distance  was defined 
\begin{equation} 
d_c([\Ybf], [\Zbf]) =  \sqrt{\sum_{i=1}^p \sin^2 \theta_i} 
\end{equation}
which is locally-equivalent to~\eqref{eq:dgG} as $\sin^2 \theta_i \approx \theta_i^2$ when $\theta_i \approx 0$, and corresponds to a Euclidean embedding since 
\begin{eqnarray} 
d_c([\Ybf], [\Zbf]) &=&  \sqrt{ p- \sum_{i=1}^p \cos^2 \theta_i}  = \sqrt{ p - \|\Zbf^H \Ybf  \|_F^2 } \nonumber \\
 &=&   \frac{1}{\sqrt{2}}\|\Ybf \Ybf^H -  \Zbf \Zbf^H  \|_F \label{GrassChorDistCos} .
\end{eqnarray}
This is in fact an isometric embedding into  $\Scal^{n^2-2}({ \scriptscriptstyle \sqrt{ \frac{p(n-p)}{2n}}})$~\cite{conway-1996-5}  which  follows from mapping any $[\Ybf]\in \grass{C}{n}{p}$ to the space of detraced Hermitian matrices as 
\begin{equation}  \label{eq:DetracedProj}
[\Ybf] \to  \bar{\Pi}_{\Ybf} = \frac{1}{\sqrt{2}}\left(\Ybf \Ybf^H -\frac{p}{n}\Ibf\right),
\end{equation}
and the chordal distance is the ambient Euclidean distance
\begin{equation} \label{eq:ChordDistGrassDetrProj}
d_c([\Ybf], [\Zbf]) = \|\bar{\Pi}_{\Ybf} - \bar{\Pi}_{\Zbf} \|_F^2. 
\end{equation}
In this ambient space of dimension $n^2-1$, it can be further verified that the distance of any $\bar{\Pi}_{\Ybf}$ from the origin is $\|\bar{\Pi}_{\Ybf} - \bm{0}\|_F^2 =  \frac{p(n-p)}{2n}$, and thus $\bar{\Pi}_{\Ybf}$ belongs to a sphere of radius $\sqrt{\frac{p(n-p)}{2n}}$. 
When there is no ambiguity on the considered space, we will simply write $d_c(\Ybf, \Zbf)$, which is well-defined as the distance does not depend on the Stiefel representatives.

Alternatively, the Grassmann manifold can be expressed as the quotient space $ \grass{C}{n}{p} \cong \Ucal_n / (\Ucal_p \times \Ucal_{n-p}) \subset \mathbb{C}^{n \times n}$. In this representation, elements in a Grassmann manifold are equivalence classes of unitary matrices. 
The tangents of $\grass{C}{n}{p} $ at the identity  are  of the form
$\Deltabf_* = \left(	\begin{array}{cc} 0  & -\Bbf^H \\ \Bbf &0
\end{array} \right)$ for this second quotient representation. Again,
we do not consider this representation, but we note here that the
natural  geodesic distance induced by this embedding only differs  
from (\ref{eq:dgG})
by a scaling factor,
\begin{eqnarray} 
d_{g*}([\Ybf], [\Zbf]) &=& \|\Deltabf_*  \|_F  = \sqrt{2} \|\Bbf  \|_F \nonumber \\
  &=&\sqrt{2} d_g([\Ybf], [\Zbf]).
\end{eqnarray}


\section{Packing Problem and Maximum Code Density}
A packing is a maximal set of non-intersecting balls of fixed radius,
covering the space so that it is not possible to fit in
another ball. For a given code size, a packing thus gives the maximum
density, i.e. the maximum fraction of the space that one can cover by
non-intersecting balls. This problem is considered to be the dual of a
coding problem: maximizing the code cardinality for a given minimum
distance or reciprocally maximizing the minimum distance for a given
code cardinality. Surprisingly, these two problems are not necessarily
equivalent for the Grassmann and Stiefel manifolds with chordal
distance. This is because there is not necessarily a one-to-one
mapping between the minimum distance and the kissing radius of the
code, which follows from the choice of an extrinsic distance combined
with the  
two-point \emph{in}homogeneity of the
spaces~\cite{Wang52,Bachoc08}.

\subsection{Code, Minimum Distance, and Metric Balls}

An $(N,\delta)$-code is a finite subset of $N$ elements in $\Mcal$,
\begin{equation}\Ccal= \{\Cbf_1,\ldots, \Cbf_N \} \subset \Mcal, \end{equation}
  where $\delta$ is the minimum distance   defined as 
\begin{equation}\label{eq:MDdef}
\delta=\min\{d_c(\Cbf_{i},\Cbf_{j})~;~ \Cbf_{i}, \Cbf_{j} \in \Ccal, i\neq j\}.
\end{equation}

A metric ball $B_{\Cbf}(\gamma) \subset \Mcal $   of radius $\gamma$ with center $\Cbf \in \Ccal$ is the subset
\begin{equation} 
B_{\scriptscriptstyle \Cbf }(\gamma) = \left\{ \Pbf \in \Mcal \; \big| \;   d_c( \Pbf,\Cbf) \leq \gamma \right\} \subseteq  \Mcal .
\end{equation}

\subsection{Kissing Radius} 
Given a code,  one can surround each codeword by metric balls of the same radius and enlarge them until two balls touch. This leads to the notion of \emph{kissing radius}. The kissing radius is a generalization of the so-called \emph{packing radius}.  We choose a different terminology to emphasize here that the kissing radius of a code may not be a function of the minimum distance, and also because the packing radius is sometimes defined as $\delta/2$ irrelevantly of the choice of distance~\cite{dhillon-2007}  by extension from flat geometry. 
\begin{Def} 
The  \emph{kissing radius} of a code $\Ccal$ is the maximum radius of non-overlapping
metric balls centered at the codewords:
\begin{equation} 
\varrho = \sup_{ \substack{ B_{\Cbf_k}(\gamma) \cap B_{\Cbf_l}(\gamma)= \emptyset \\ \Cbf_k \neq \Cbf_l }}{\gamma}.
\end{equation}
\end{Def}

For each codeword pair $(\Cbf_k,\, \Cbf_l)$,  there exists a mid-point $\Mbf_{k,l}$ which is the closest equidistant point satisfying $\Mbf_{k,l} = \arg \min_{\Mbf} d_c(\Cbf_k,\Mbf)$ subject to $d_c(\Cbf_k,\Mbf) = d_c(\Cbf_l,\Mbf) $. The mid-points $\Mbf_{k,l}$ define the mid-distances $\varrho_{k,l} = d_c(\Cbf_k,\Mbf_{k,l}) = d_c(\Cbf_l,\Mbf_{k,l}) $, and the kissing radius is the minimum of all of them  $\varrho = \min_{k\neq l}{\varrho_{k,l}}$. 

As the chordal distance is measured along a geodesic defined by the associated Riemann metric, the midpoints can be computed accordingly: 
with $\Cbf_{k,l}(t)$ the geodesic  such that $\Cbf_{k,l}(0) = \Cbf_k $ and $\Cbf_{k,l}(1)= \Cbf_l$, the midpoint is  $\Mbf_{k,l}= \Cbf_{k,l}(1/2)$. 
In the case of the unitary group and  the  Grassmann manifold, the geodesic equation is fully parametrized by a set of principal   
angles $\{\theta_i \}$ as given in~\eqref{eq:UnEigDecomp} and~\eqref{eq:GrassGeodesicEq}, respectively, and the mid-distance $\varrho_{k,l}$  can then be directly computed from half of these angles  $\{\frac{\theta_i}{2} \}$. 
However, for the case of the Stiefel manifold, the geodesic equation between two points is not known as explained in Remark~\ref{Rem:RiemLogMap}. Alternatively, the midpoint can  be computed\footnote{Except if $\Cbf_k$ and  $\Cbf_l$ are antipodal then their center of mass is $\bm{0}$ which does not have a unique projection.} by an orthogonal projection of the center of mass $\frac{\Cbf_k + \Cbf_l}{2}$, which for the Stiefel manifold is given by polar decomposition~\cite{TWC14}. 
 
For spherical codes one has only one principal angle. This results in a one-to-one mapping between the minimum distance and kissing
radius. In contrast, for the Grassmann and Stiefel manifolds with
chordal distance, the kissing radius and the minimum distance may not
be directly expressible as functions of each other.  
First, the kissing radius corresponds to the mid-distance between two
codewords, which may  
not be at the minimum distance from each other.  
Second, for a given pairwise distance, one may have different
mid-distances. Grassmann and Stiefel manifolds are not in
general two-point homogeneous spaces~\cite{Wang52,Bachoc08}: one
cannot necessarily find a unitary mapping between two pairs of
equidistant points, i.e., a pair $(\Cbf_k, \Cbf_l)$ cannot always be
mapped to a pair $(\Cbf_i, \Cbf_j)$ even if $d_c(\Cbf_k, \Cbf_l) =
d_c(\Cbf_i, \Cbf_j) $. In fact, in the case of $\Mcal = \Ucal_n$ and
$\grass{C}{n}{p}$, the collection of principal angles provides the
complete relative position between two points which is transitive
under the action of the unitary group. However, by compressing this
``vector-like distance'' to the scalar distance $d_c$, one loses
transitivity.

\subsection{Density}

The density of a  code is the fraction of $\Mcal$ covered by metric balls centered around the codewords with radius equal to the kissing radius. 
We consider a uniform measure $\mu$ on $\Mcal$ inherited from the Haar measure on the unitary group. Recall that the Grassmann and Stiefel manifolds are homogeneous spaces of the unitary group. For any measurable set $\mathcal{S}\subset \Mcal$ and any $\Ubf \in \Ucal_n$, the uniform measure satisfies $\mu\left(\Ubf\mathcal{S}\right)=\mu\left(\mathcal{S}\right)$. Due to the homogeneity of $\Mcal$, the characteristics of this ball are independent of its center which for convenience will often not be specified. 
The measure $\mu$ corresponds to a normalized volume 
\begin{equation}  \label{eq:NormalizeVol}
\mu(B(\gamma)) = \frac{\text{vol } B(\gamma)}{\text{vol } \Mcal},
\end{equation} 
satisfying $\mu(\Mcal)=1$.

\begin{Def} The density of a code $\Ccal \in \Mcal$ is defined as 
\begin{eqnarray}  
\Delta(\Ccal) &=&  \mu \left(\bigcup_{\Cbf_i \in\Ccal } B_{\Cbf_i}(\varrho) \right) =\sum_{\Cbf_i \in\Ccal }\mu \left(B_{\Cbf_i}(\varrho) \right)  \nonumber \\
 &=& N \mu \left( B(\varrho) \right). 
\end{eqnarray}
\end{Def}

\begin{table*}
\caption{An example maximal-minimum-distance code   $\mathcal{C}_1$ in $\grass{C}{4}{2} $   which is not an optimal packing since it has a lower density than $\mathcal{C}_2 $. \label{tab:exampleCodeG42}}
\vspace{-0.4cm}
\begin{center}
\begin{tabular}{|c|c|}	
 \hline	
$\mathcal{C}_1$   & $\mathcal{C}_2$
 \\\hline
density $\Delta = \frac{8}{9} \left(7-4 \sqrt{3}\right)\approx 0.0638$  
& density $\Delta =\frac{1}{8} = 0.125$  \\
min. dist. $\delta = \frac{2}{\sqrt{3}} \approx 1.15$ & min. dist. $\delta = 1$ \\ 
kissing radius $\varrho = \sqrt{2} \alpha_{-}\approx0.65$ 
&
kissing radius $\varrho =\frac{1}{\sqrt{2}} \approx 0.71$ 
\\\hline
  $  \left\{ \begin{bmatrix} \alpha_{+} & 0  \\ \alpha_{-} & 0 \\  0 & \alpha_{+}  \\ 0 &\alpha_{-}   \end{bmatrix} \;	
	    \begin{bmatrix} \alpha_{+} & 0  \\ -\alpha_{-} & 0 \\  0 & \alpha_{+}  \\ 0 &-\alpha_{-}   \end{bmatrix} \;	
		 \begin{bmatrix}\alpha_{-}  & 0  \\  i \alpha_{+} & 0 \\  0 &\alpha_{-}   \\ 0 & i \alpha_{+}   \end{bmatrix} \;
		\begin{bmatrix}\alpha_{-}  & 0  \\  -i \alpha_{+} & 0 \\  0 &\alpha_{-}   \\ 0 & -i \alpha_{+}   \end{bmatrix} \; \right\} $  
&
$  \left\{ \begin{bmatrix} 1 & 0  \\ 0& 1 \\  0 & 0 \\ 0 &0  \end{bmatrix} \;	
	 \begin{bmatrix}  0 & 0 \\ 1 & 0  \\ 0& 1 \\  0 &0  \end{bmatrix} \;	
			 \begin{bmatrix} 0 & 0 \\ 0 &0 \\ 1 & 0  \\ 0& 1 \end{bmatrix} \;	
			 \begin{bmatrix}  0& 1 \\  0 & 0 \\ 0 &0 \\ 1 & 0  \end{bmatrix} \;	\right\} $ 
		\\
	where 	$\scriptstyle \alpha_{\pm}=\sqrt{\frac{1}{6} \left( 3 \pm \sqrt{3} \right) } $ &  \\\hline
\end{tabular} 
\end{center}
\end{table*}

\begin{table}
\caption{Infinite code families $\mathcal{C}^m_1$ and  $\mathcal{C}^m_2$ in  $ \grass{C}{2^m}{2} $ generalizing the codes in Example~\ref{Example1} \label{tab:exampleCodesTensor}}
\vspace{-0.4cm}
\begin{center}
\begin{tabular}{|c|c|} \hline
$\mathcal{C}^m_1$   & $\mathcal{C}^m_2$
 \\\hline
Cardinality $N = 4^{m-1}$ & Cardinality $N = \binom{2^{m}}{2} $ \\ 
min. dist. $\delta = \frac{2}{\sqrt{3}} \approx 1.15$ & min. dist. $\delta = 1$ \\ 
kissing radius $\varrho = \sqrt{2} \alpha_{-}\approx0.65$ 
&
kissing radius $\varrho =\frac{1}{\sqrt{2}} \approx 0.71$ \\ \hline
\end{tabular} 
\end{center}
\end{table}

By definition, it satisfies  $\Delta(\Ccal) \leq 1$. 
A maximum packing code has the maximum density for a given cardinality.   
The problem of maximizing the density for a given cardinality corresponds to maximizing the kissing radius of the code. This could be in fact reformulated as another maximum-minimum distance problem where distances among codewords would be defined by the mid-distances. In the case of the Grassmann manifold and unitary group, this would correspond to halve the principal angles in the chordal distance definition, which for the Grassmann manifold can be  identified  as the ``chordal Frobenius-norm''~\cite{Edelman} (up to a scaling factor). 
 
In general, the packing problem is not necessarily equivalent to maximizing the minimum distance of the code, and codes with same minimum distance may also have different kissing radiuses and thus densities. 
We illustrate these statements with two concrete examples below.   
The first example describes two  
Grassmann codes  
where the (proven) maximum minimum-distance code has a lower density than another code.

\begin{Example} \label{Example1}
Consider the two Grassmannian codes given in
Table~\ref{tab:exampleCodeG42}. The density of these codes of
cardinality four in $\grass{C}{4}{2}$ is $\Delta = 2 \varrho^8$
according to the volume formula provided in~\cite{Dai08}.

The first code $\Ccal_1$ is an optimal max-min distance code reaching
the Rankin simplex bound~\cite{Rankin,conway-1996-5}, constructed by
embedding the optimal tetrahedron code of
$\grass{C}{2}{1}$~\cite{TIT11}. The embedding is obtained by a tensor
product with the identity matrix,  
following~\cite[Prop.~12]{CreignouarXiv}. This is a strongly simplicial
configuration in the sense that all principal angles equal
$\arccos \frac{1}{\sqrt{3}}$. The mid-distances between each  
pair of
codewords are thus all the same, approximately $0.65$, leading to a
density of $\approx 0.0638$.

The second code $\Ccal_2$ is obtained by circular permutation of the rows of the truncated identity matrix. Its distance distribution corresponds to the embedding of a square.  
The principal angles between two codewords are either $\{0,\frac{\pi}{2} \}$  or $\{\frac{\pi}{2},\frac{\pi}{2} \}$. The mid-distances between each of the codewords are either $1/\sqrt{2}$ or $1$. 
The density of this code is $0.125$, almost twice the density of the optimum max-min distance code $\Ccal_1$ while it has a smaller minimum distance.
\end{Example}

The codes in Example~\ref{Example1} can be generalized to two infinite families of codes $\mathcal{C}^m_1$ and  $\mathcal{C}^m_2$ in  $\grass{C}{2^m}{2}$ where both have constant minimum distance and kissing radius:   
Given $m\geq 2$, the codewords   $ \Cbf_k \in \mathcal{C}^m_1$   are
constructed by the   
$m$-fold tensor product $ \Cbf_k = \Ibf_2 \otimes \cbf_1 \otimes \cdots \otimes
\cbf_{m-1}  $ for all $ \cbf_i \in \left\{ [ \alpha_{+},
  \pm \alpha_{-}]^T , \; [ \alpha_{-}, \pm i \alpha_{+}]^T  \right\}
  $, 
leading to a code with $4^{m-1}$ codewords in $\grass{C}{2^{m}}{2}$ with constant minimum distance $\delta = 2/\sqrt{3}$ and kissing radius $\varrho = \sqrt{2} \alpha_{-}\approx0.65$. 
The code  $\mathcal{C}^m_2$ is obtained by row permutations of the 
truncated identity matrix $\Ibf_{2^m,2}$ leading to $\binom{2^{m}}{2}$
codewords with constant minimum distance $\delta =1$ and kissing
radius $\varrho =1/\sqrt{2} \approx 0.71$. For all $m$, the first
construction $\mathcal{C}^m_1$ has a larger minimum distance than
$\mathcal{C}^m_2$, but the latter has a larger kissing radius and
density as summarized in Table~\ref{tab:exampleCodesTensor}. The code
$\mathcal{C}^m_2$ is  
larger than $\mathcal{C}^m_1$ but one could select only $4^{m-1}$
codewords as in $\mathcal{C}^m_1$ without changing the kissing radius
and minimum distance.

\begin{table*}
\caption{Example of two maximal-minimum-distance codes $\mathcal{C}_3$ and $\mathcal{C}_4$ in $ \grass{C}{7}{3}$ with different densities~\cite{Calderbank99agroup-theoretic}. \label{tab:exampleCodeG73}}
\vspace{-0.4cm}
\begin{center}
\begin{tabular}{|c|c|}	
 \hline	
$\mathcal{C}_3$  & $\mathcal{C}_4$  
 \\\hline
density $\Delta \approx 10^{-4.2}$ & density $\Delta \approx 10^{-3.5}$    \\ 
min. dist. $\delta = \frac{4}{3}$& min. dist. $\delta = \frac{4}{3}$ \\
kissing radius  $\varrho \approx 0.75$  & kissing radius  $\varrho \approx 0.81$ 
\\\hline
Generator matrices:   & Generator matrices: \\  
	 $  \begin{bmatrix} 0 & 1 & 0 & \pm \sqrt{2} & 0 & 0 & 0  \\
	                 0 & 0 & 1 & 0 & 0 & 0 &  \pm \sqrt{2}  \\
									0 & 0 & 0 & 0 & 1 & \pm \sqrt{2} & 0  \\
									\end{bmatrix}^T $
&	       $  \begin{bmatrix} 0 & 1 & 0 & 0 & 0 & 0 & \pm \sqrt{2}   \\
	                 0 & 0 & 1 & 0 & 0 & \pm \sqrt{2} &  0  \\
									0 & 0 & 0 & \pm \sqrt{2}  & 1 & 0 & 0  \\
									\end{bmatrix}^T \;	
	     $  
	\\\hline
	\multicolumn{2}{|c|}{The  signs are selected such that their product is $+1$. } 
	\\  
	\multicolumn{2}{|c|}{28 codewords  obtained by  circular  permutation of  the rows of the four generator matrices}
	\\\hline
\end{tabular} 
\end{center}
\end{table*}

The second example shows that two optimal maximal-minimum-distance codes with  different densities can exist. 
 
\begin{Example} \label{Example2}
In~\cite{Calderbank99agroup-theoretic}, an infinite family of real Grassmann codes meeting the Rankin simplex bound is described generalizing a code found in~\cite{conway-1996-5}. Since the simplex bound is the same for real and complex Grassmannians, these codes are also maximum-minimum-distance complex Grassmann codes. It is observed that this construction can lead to different codes with same minimum distance but a different distribution of principal angles, and thus different densities. This observation follows from the lowest dimensional examples  $\mathcal{C}_3$ and $\mathcal{C}_4$ in Table~\ref{tab:exampleCodeG73}. 
Each code is the union of four orbits under the action of the cyclic group but from different generator matrices.   
The density of these codes of cardinality 28 in $\grass{C}{7}{3}$ can
be approximated~\cite{Dai08} to $\Delta \approx \frac{2}{33}
\varrho^{24}$. 
The code $\mathcal{C}_3$ has three distinct sets of principal angles and thus three distinct mid-distances from which the kissing radius is  $\varrho = \sqrt{\frac{9-\sqrt{2}-\sqrt{3}-\sqrt{6}}{6}} \approx 0.75$. The code $\mathcal{C}_4$ has two distinct sets of principal angles and thus two mid-distances, among which the set of angles  $\{  0, \arccos \frac{1}{3}, \arccos \frac{1}{3} \}$ with mid-distance $\sqrt{2/3}\approx 0.817$ is also a mid-distance of $\mathcal{C}_3$. The kissing radius is achieved with the other set of principal angles which has a mid-distance only slightly smaller $\approx 0.805$ (it does not seem to have a compact form). 
While both codes reach the optimal minimum distance of
$\frac{4}{3}$, the code $\mathcal{C}_4$ has a density about 5 times
larger than $\mathcal{C}_3$.

\end{Example}

We briefly  discuss the performance of these codes when applied to MIMO communications.  

MIMO precoding is well-known to be related to Grassmannian
packing~\cite{overview}. In this context, the mutual information with
Gaussian signaling is given by   
$ {\mathbb E}_{\Hbf}\left[ \log_2 \det \left(\Ibf + \gamma
  \Cbf_{\rmq(\Hbf)}^H \Hbf^H \Hbf\Cbf_{\rmq(\Hbf)} \right)\right] $
where $\gamma$ is the per-stream SNR, and
$\rmq(\cdot)$ is a quantization map that selects the codeword index in
$\Ccal$ maximizing the instantaneous rate for each channel realization
$\Hbf$ inside the expectation. Entries of $\Hbf \in \Cbb^{ p \times
  n}$ are assumed to be standard complex normal variables
$\Ccal\Ncal(0,1)$.  
Numerical evaluations of corresponding mutual informations for the codes in Examples~\ref{Example1} and~\ref{Example2} show that the two higher density codes $\Ccal_2$ and $\Ccal_4$ slightly outperform the codes $\Ccal_1$ and $\Ccal_3$, respectively.  $\Ccal_2$ provides a  SNR gain of  $0.16$dB over  $\Ccal_1$, and   $\Ccal_4$ provides a  SNR gain of $0.05$dB over  $\Ccal_3$, in the SNR region $p\gamma = 20$dB. 
These examples and other numerical experiments hint that it may be in fact the density of the code that primarily governs the achievable rate of transmission rather than the maximum-minimum distance. We believe this behavior to be quite generic for the following reason. 
MIMO precoding reduces to a quantization problem on the Grassmann
manifold with chordal distance~\cite{Dai08}, and the mid-distances of
the code reflect better the rate-distortion trade-off since they
represent the first border effects of the Voronoi cells.

\begin{figure}
\begin{center}
  \includegraphics[width=0.5\textwidth]{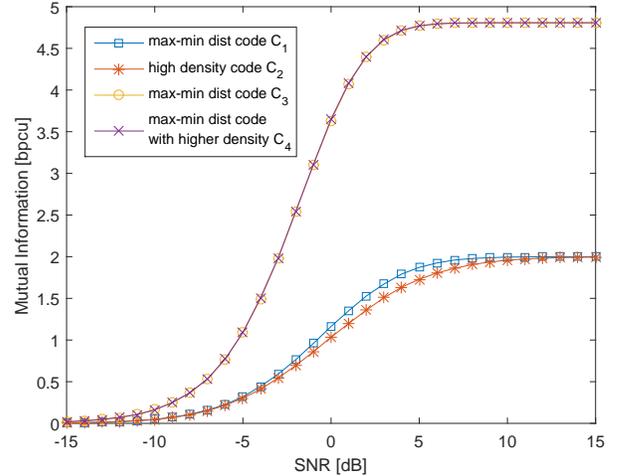}
\end{center}
\vspace{-0.5cm}
  \caption{Comparison of the codes in Examples~1 and~2 when used as a
    space-time constellation  
for non-coherent communication.  
 \label{fig:MutInf_ExampleCBs}
}
\end{figure}

We now look at the application of unitary codes as space-time
constellations over a non-coherent MIMO transmission  $ \Ybf =
\sqrt{{\rm snr} \frac{n}{p}} \Cbf \Hbf + \Nbf $. 
Here $\Hbf\in \Cbb^{p \times p}$ and $\Nbf\in \Cbb^{n \times p}$ have
standard complex normal $\Ccal\Ncal(0,1)$ entries, and $\Cbf$ is
uniformly selected from the code $\Ccal$. The standard approach in
this case is to design Grassmann codes that maximize the product
diversity $\prod_i \sin^2 \theta_i$  
since it is minimizing the pairwise block error rate at
high-SNR~\cite{HochwaldMarzettaTIT00}. Low-SNR
analysis~\cite{HanRosenthalGeo} shows on the other hand that it is the
chordal distance (sum diversity) $\sum_i \sin^2 \theta_i$ that
dominates the block error rate, and it was additionally observed
in~\cite{AshikhminCalderbankTIT10} that codes maximizing the chordal
distance can lead to higher mutual information $I(\Ybf;\Cbf)$ than
codes with high product diversity. Furthermore, 
subspace perturbation analysis in~\cite{GoharyTIT09} suggests 
that an appropriate code metric is given by the chordal
Frobenius-norm, which is exactly the kissing radius of the code with
chordal distance as explained above, and thus its density. 
From the evaluation of the mutual informations $I(\Ybf;\Cbf)$ of the
codes in Examples~\ref{Example1} and~\ref{Example2} shown on
Figure~\ref{fig:MutInf_ExampleCBs}, the code density does not  
seem to be an ultimate  performance measure  
when it comes to mutual information, 
as $\Ccal_1$ clearly outperforms $\Ccal_2$. It should be noted that $\Ccal_1$ has both a higher minimum chordal distance and product diversity than $\Ccal_2$, the later having zero product diversity between four codeword pairs out of six. 
 
\subsection{Hamming Bound}
The notion of density directly relates to the Hamming bound:   
For any $(N,\delta)$-code, one must satisfy 
\begin{equation}  
\label{bd:hamStd}
N \leq \frac{1}{\mu(B(\frac{\delta}{2}))} .
\end{equation} 
The upper bound in~\eqref{bd:hamStd} is a direct application of the standard Hamming bound to  $\Mcal$ without taking into account the curvature of the space and the choice of distance. 
The chordal distance,  inherited from a Euclidean embedding, is extrinsic to the considered curved space and thus never satisfies the triangle inequality with equality. 
Accordingly, balls of radius $\delta/2$ around the codewords  do not necessarily form a packing, as none of the balls would be touching each other and one could possibly fit in an extra ball. The kissing radius $\varrho $ is larger than $\delta/2$, and the Hamming bound can be refined by any radius $\delta/2 < r \leq \varrho $, and ultimately for $r = \varrho$,
\begin{equation}\label{eq:ImpHB}
N \leq \frac{1}{\mu(B(\varrho))}. 
\end{equation}
By construction  $ \frac{\delta}{2} \leq \varrho $ and thus the Hamming bound~\eqref{eq:ImpHB} is always tighter than the `standard Hamming bound'~\eqref{bd:hamStd}, while being asymptotically equivalent for $\delta \to 0$.

The difficulty in exploiting the improved bound~\eqref{eq:ImpHB} is   
in finding a relationship between the kissing radius and the minimum
distance of the code. To obtain a bound on the minimum distance, one
needs to find a function of the minimum distance such that $ \delta/2
\leq f(\delta) \leq \varrho $. Then, provided that both the volume
expression and $f$   
are invertible, it  
is 
possible to bound the minimum distance from above.

\section{Volumes}
In this section, we address the problem of volume computation in the manifolds, providing two different asymptotic approximations of the volume of balls operating in different regimes.

\subsection{Spherical Volumes and Hyperspherical Caps}
The $(D-1)$- and $D$-dimensional volume of $\Scal^{D-1}(R)$ (with its natural metric $d_c$) are respectively
\begin{equation}
\label{eq:vol:sph}
A_D(R) = \frac{2 \pi^{D/2}}{\Gamma\left(\frac{D}{2} \right)} R^{D-1}, \quad V_D(R) = \frac{\pi^{D/2}}{\Gamma\left(\frac{D}{2}+1 \right)} R^{D} .
\end{equation}

Since the manifolds of interest are submanifolds of hyperspheres, the considered balls are subsets of hyperspherical caps. 
A hyperspherical cap is a ball on a sphere,  
 \begin{equation}
C_{D,R}(r)= \left\{ \xbf \in \Scal^{D-1}(R) \; : \quad   \|\xbf-\ybf\| \leq r\right\} 
\end{equation}
for some implicit center $\ybf$.   
One can define a uniform spherical measure $\sigma$,  
and the  normalized volume of the spherical caps is denoted by $\sigma(C_{D,R}(r))$. 
We use a different notation for distinction with the uniform measure on the manifold $\Mcal \subset \Scal^{D-1}(R)$ but if considering $\Mcal=\Scal^{D-1}(R)$ then the two measures match $\mu (B(r)) = \sigma(C_{D,R}(r)) $. 

The $(D-1)$-dimensional volume of a spherical cap can be computed exactly, as given below. It is given along with two asymptotics proven in Appendix~\ref{proofs:Cap}.  
\begin{Lem} \label{Lem:vol_caps} 
The normalized volume (area) of a hyperspherical cap in $\Scal^{D-1}(R)$ measured with chordal distance is given by 
\begin{equation}
\label{eq:vol_caps}
 \sigma(C_{D,R}(r)) = 	I_{\frac{r^2}{4R^2}}\left(\frac{D-1}{2}, \frac{D-1}{2} \right)  
\end{equation}
where  $r$ is the radius of the cap satisfying $0 \leq r \leq 2 R$, and $I_{x}(a,b)$ is the regularized incomplete beta function.

As the  radius of the cap $r$ goes to zero, the volume  of the spherical cap tends to  
\begin{equation} \label{eq:smallcap}
\sigma(C_{D,R}(r)) \simeq   \frac{1}{2 \sqrt{\pi}} \frac{\Gamma(\frac{D}{2})}{\Gamma(\frac{D+1}{2})} \left(\frac{r}{R}\right)^{D-1}  .
\end{equation}
On the other hand, as the dimension of the sphere $ D$ goes to infinity with $\sqrt{D}  \left(1-\frac{r^2}{2R^2} \right )$ fixed, the volume tends to
\begin{equation} \label{eq:largecap}
 \sigma(C_{D,R}(r))  \simeq \frac12 \erf\left( \sqrt{\textstyle \frac{D}{2}}\right) - \frac12 \erf \left( \sqrt{ \textstyle\frac{D}{2}}  \left(1-\frac{r^2}{2R^2} \right )\right)   
\end{equation}
where $\erf(x)$ is the Gauss error function defined in Eq.~\eqref{eq:erf}. 
\end{Lem}

The volume~\eqref{eq:smallcap} equals $V_{D-1}(r)/A_{D}(R)$. Intuitively, when a cap is small it is almost ``flat'' and its volume tends to be the ``area of a disc''. 
The volume~\eqref{eq:largecap} corresponds to an asymptotic Gaussian behavior for large dimensions: 
the squared chordal distance of uniformly distributed random points is asymptotically Gaussian; $d_c^2 \sim \Ncal\left(2R^2, \frac{4R^4}{D} \right)$ where $ \Ncal(m,v)$ is the normal distribution with mean $m$ and variance $v$. A similar expression formulating this high-dimensional regime can be found in~\cite{Chudnov1,Chudnov2}. 
As $D\to \infty$, high-dimensional random vectors are asymptotically orthogonal and the chordal distance tends to $\sqrt{2}R$ which is the mid-distance between two antipodal points.

\begin{figure*}
\begin{center}
 \subfigure[Unitary group $\Ucal_n$]{\includegraphics[width=0.5\textwidth]{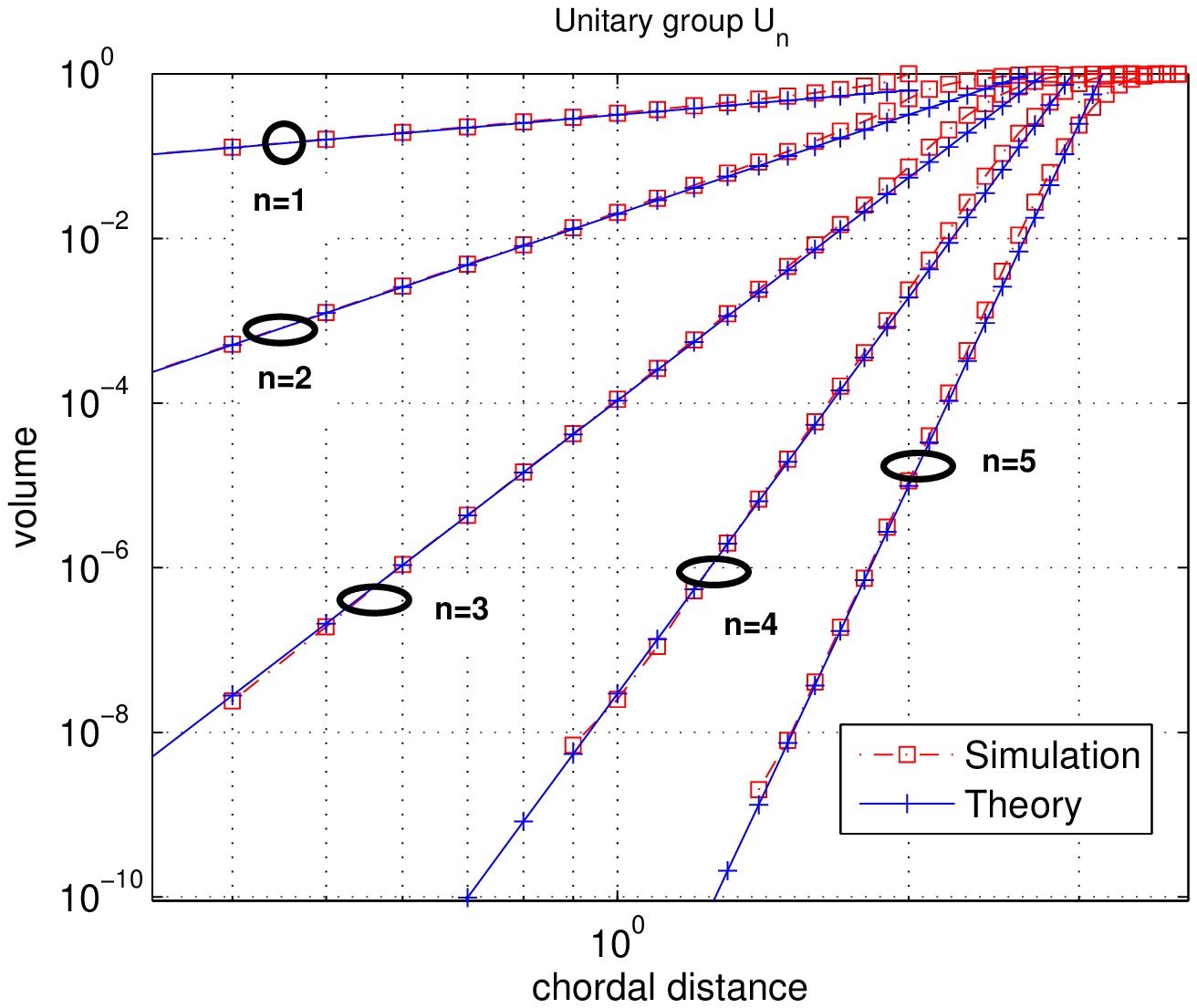}}
 \subfigure[Stiefel manifold $\stief{C}{n}{p}$]{\includegraphics[width=0.49\textwidth]{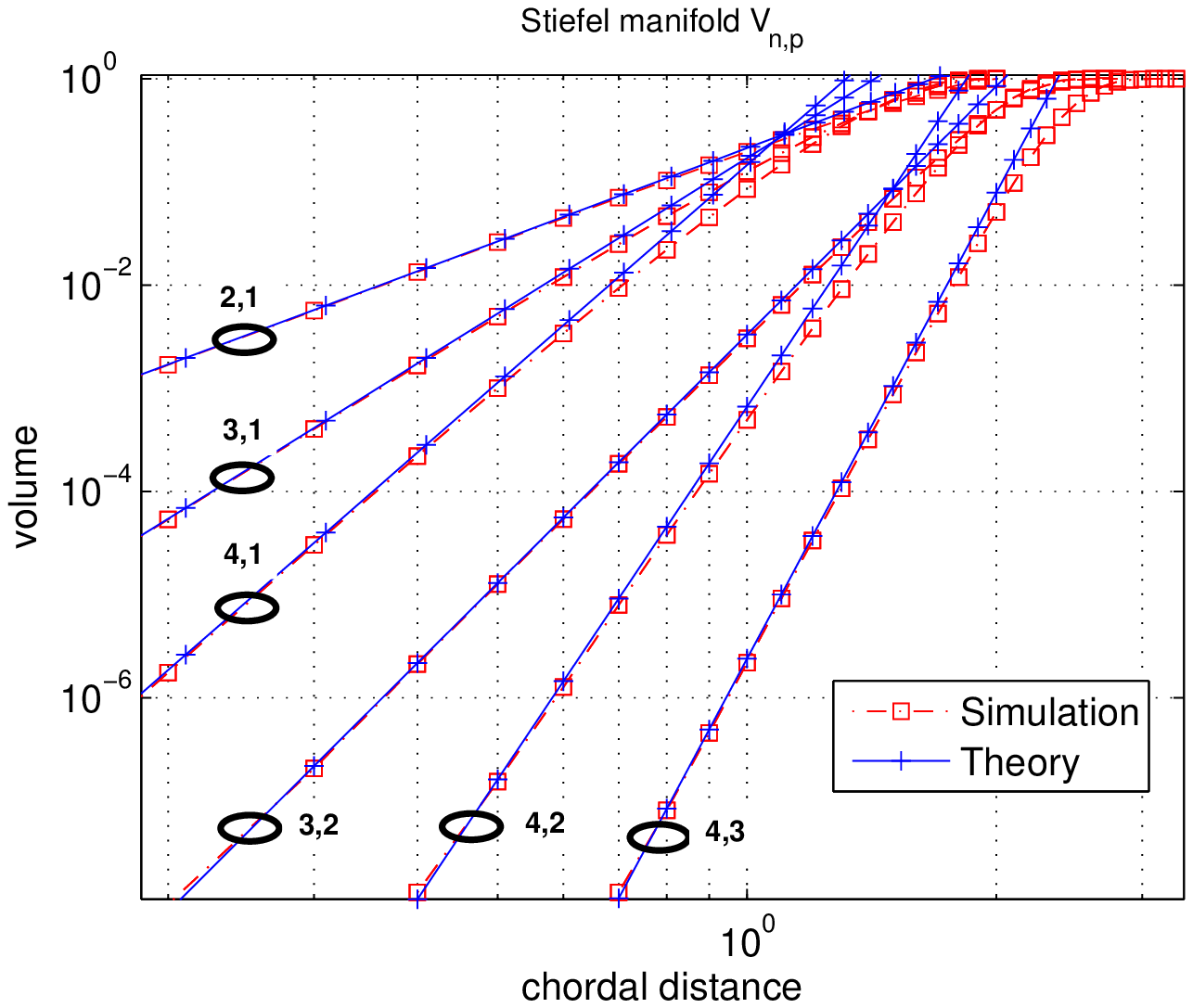}}
\end{center}
\vspace{-0.3cm}
 \caption{The small ball volume approximation~\eqref{eq:SmallApprox} of Corollary~\ref{cor:vol} compared to simulation. 
 \label{fig:vol_un}
}
\end{figure*}

\subsection{Manifold Volume and Small Ball Approximation}
\subsubsection{Overall Manifold Volume}
The volume of a space can be obtained from the integration of a volume
element, 
which is unique on a Riemannian manifold up to a non-vanishing scaling
factor. This scaling factor is induced by the chosen distance, and
impacts the overall volume. The theorem below provides the volumes
correctly scaled for the chordal distance. 
\begin{Thm}
\label{thm_vol_stief}
The volumes of the manifolds induced by the chordal distance $d_c$ (or its equivalent geodesic distance $d_g$) are
\begin{itemize}
\item for the unitary group~\cite{Hua}
\begin{equation}
{\rm vol }\; \Ucal_n  =   \frac{(2\pi)^{\frac{n(n+1)}{2}} }{\prod_{i=1}^n (i-1)!},
\end{equation}

\item  for the Stiefel manifold

\begin{equation}
{\rm vol }\; \stief{C}{n}{p}  =   \frac{2^{\frac{p(p+1)}{2}} \pi^{np - \frac{p(p-1)}{2}}}{ \prod_{i=1}^p (n-i)!
},
\end{equation}

\item for the Grassmann manifold
\begin{equation} 
{\rm vol }\;  \grass{C}{n}{p} = \pi^{p(n-p)} \prod_{i=1}^p \frac{(p-i)!}{(n-i)!}
\end{equation}
\end{itemize}
\end{Thm}
A detailed volume computation is provided in Appendix~\ref{proof:overall_volume}.  
While it is possible to find the exact volume for the unitary group rigorously derived for the chordal distance in~\cite{Hua,Zyczkowski}, the equivalent result for the Stiefel manifold does not seem to have been reported before.   
This volume differs by a constant factor from the commonly cited formula, see e.g.~\cite{Tse,Barg,henkel}. In many contexts, the overall scaling of the volume is meaningless as often the induced scaling  would be absorbed or canceled out.   The volumes known in the literature arise from  integration of a volume element neglecting scalar prefactors. Conventional volumes in the literature are  expressed as the product of the volumes of spheres:  $\displaystyle \mathfrak{vol}  \; \stief{C}{n}{p} = \prod_{k=n-p+1}^{n} V_{2k}(1)= \prod_{k=n-p+1}^{n} \frac{2 \pi^k}{(k-1)!},$ where $ V_{2k}(1) = \text{vol } \Scal^{2k-1} $ is given in~\eqref{eq:vol:sph}. 
The two different conventions are related by ${\rm vol }\; \stief{C}{n}{p} = 2^{\frac{p(p-1)}{2}} \mathfrak{vol}  \; \stief{C}{n}{p} $. 
For the Grassmann manifold  with the geometry induced by the spherical embedding, it appears that these normalizations cancel out so that we have 
\mbox{$ \displaystyle  {\rm vol } \;\grass{C}{n}{p}  = \frac{{\rm vol } \; \stief{C}{n}{p} }{ {\rm vol }\; \Ucal_p } = \frac{\mathfrak{vol } \; \stief{C}{n}{p} }{ \mathfrak{vol }\; \Ucal_p } = \frac{\mathfrak{vol } \; \Ucal_n  }{ \mathfrak{vol }\; \Ucal_p  \mathfrak{vol }\; \Ucal_{n-p} } $ }. 
However, we remark that the true volume of the Grassmannian with the  quotient geometry $\grass{C}{n}{p} \cong \Ucal_n/(\Ucal_p \times \Ucal_{n-p} )$ and distance $d_{g*}$ is 
\mbox{$\displaystyle   {\rm vol_* } \;\grass{C}{n}{p}  = \frac{{\rm vol } \; \Ucal_n  }{ {\rm vol }\; \Ucal_p  {\rm vol }\; \Ucal_{n-p} } $}, 
which differs from  $ {\rm vol } \;\grass{C}{n}{p} $ in the embedding geometry computed above by a factor of $2^{p(n-p)}$.  Using the conventional volumes, one would  get 
$\displaystyle  \mathfrak{vol_* } \;\grass{C}{n}{p} = \frac{\mathfrak{vol } \; \Ucal_n  }{ \mathfrak{vol }\; \Ucal_p  \mathfrak{vol }\; \Ucal_{n-p} } =  {\rm vol } \;\grass{C}{n}{p}$, 
i.e. the volumes in the two geometries would erroneously be assumed to be the same.  
When interpreting the Haar measure as a normalized volume~\eqref{eq:NormalizeVol}, keeping track of the scaling of the volume is necessary. We suspect that omitting the volume scaling factors  is one of the reasons behind the numerical errors observed in~\cite{henkel},  in which a standard volume formula was used irrespectively of the geometric interpretation of the Stiefel and Grassmann manifolds, and the related distances. 

Considering $d_c$, some special manifolds of interest are isometrically isomorphic to spheres: the Stiefel manifolds $\stief{C}{n}{1} \cong \Scal^{2n-1}$, and the Grasmann manifolds $\grass{C}{2}{1} \cong \Scal^{2}(\frac{1}{2})$. As expected, the volumes in Theorem~\ref{thm_vol_stief} match the spherical volume~\eqref{eq:vol:sph}  in these cases.

\subsubsection{Small Ball Approximation}
The volume of a small \emph{geodesic} ball can be well approximated by the volume of a ball of equal radius in the tangent space. This approximation, tight as the radius goes to zero,  is actually an upper bound  
as discussed in~\cite{henkel}, known as the Bishop--Gromov inequality, a volume comparison theorem valid for any  Riemannian manifold. In~\cite{henkel},  it was used to evaluate volumes of small geodesic balls  in the Grassmann and Stiefel manifolds. For the case of the Stiefel manifolds, results were extended to the chordal distance in an indirect manner from the geodesic distance $d_{g*}$ combined with local inequalities. 
Surprisingly, the results of~\cite{Dai08} show that this approximation
is exact for the Grassmann manifold \emph{with chordal distance}
smaller than one, in the same manner as the area of a cap of the real
sphere equals 
the area $\pi r^2$ of a disk. Later, the same approximation was used
in~\cite{love_riem} for volumes in simple flag manifolds. Refining the
result, a power series expansion for the volume of small geodesic ball
in any Riemannian manifold~\cite{Gray} was later leveraged
in~\cite{Krishnamachari_R,Krishnamachari_C}.
Limiting this expansion to the leading term gives as $r\to 0$
\begin{equation} \label{eq:GraySmallBall}
\text{vol }  B(r) = V_{\dM}(r)(1+ O(r^2))
\end{equation}
where $V_{\dM}(r)$ is according to~\eqref{eq:vol:sph} with the dimension $\dM$ of the manifold in Table~\ref{tab:emb}. Intuitively, in a small neighborhood the manifold looks like a Euclidean space and can be approximated by the tangent space. 
Other coefficients of the series expansion are addressed in~\cite{Krishnamachari_C} requiring computation of the curvature of the  manifold.

The expansion~\eqref{eq:GraySmallBall} given for the geodesic distance extends to the corresponding chordal distance induced by the isometric embedding in $\mathbb{R}^D$ as $d_g = d_c +O(d_c^3)$~\cite{Belkin_JCSS_08,BelkinPHD}. 
Therefore, the normalized volume of metric balls in $\Mcal$ with dimension $\dM$  for~\emph{both} the geodesic distance $d_g$ and the chordal distance $d_c$ is given  by
\begin{equation}
\label{eq:vol_stief}
\mu( B(r)) = \frac{V_{\dM}(r)}{ \text{vol } \Mcal }(1+ O(r^2)) 
\end{equation} 
as $r\to 0$. 
We then have the following results as a direct consequence of Theorem~\ref{thm_vol_stief} and $V_D(r)$  given in~\eqref{eq:vol:sph}.

\begin{Cor}
\label{cor:vol}
The volume of metric balls  as $r \to 0$  with metrics $d_c$ or $d_g$ is 
\begin{equation}
\mu( B(r)) = c_{n,p} r^{\dM}(1+ O(r^2)), 
\label{eq:SmallApprox}
\end{equation}
\begin{itemize} 
\item  where for  the Stiefel manifold $\stief{C}{n}{p}$, 
\begin{equation}
c_{n,p}  =  \frac{2^{-\frac{p(p+1)}{2}} \pi^{\frac{-p}{2}} }{\Gamma(p(n-p/2)+1)} \prod_{i=1}^{p} (n-i)! ,
\label{eq:cnpS}
\end{equation}
  
\item  and for the Grassmann manifold $\grass{C}{n}{p}$~\cite{Dai08},
\begin{equation}
c_{n,p}  =  \frac{1}{(p(n-p))!} \prod_{i=1}^{p} \frac{(n-i)!}{(p-i)!}.
\label{eq:cnpG}
\end{equation}
\end{itemize}
\end{Cor}

In the case of the geodesic distance $d_g$, the Bishop--Gromov inequality~\cite{henkel} is accordingly given by $\mu( B(r)) \leq c_{n,p} r^{\dM}$. 

For the Stiefel manifold with $n \neq p$, the result above 
differs from the one in~\cite{Krishnamachari_C} derived for
quotient geometry ($d_{g*}$). The two results match for  
the unitary group with $n=p$, as expected. Comparing to the
Grassmannian case in~\cite{Krishnamachari_C}, there is a difference by
a factor of $2^{2p-\frac{n-p}{2}(n-p+1)}$. The local equivalence
between the chordal distance $d_c$ and the geodesic distance $d_g$
makes the result identical for both metrics. For the Grassmann manifold with chordal distance,
the result of~\cite{Dai08} is stronger than Corollary~\ref{cor:vol} as it gives $\mu( B(r)) = c_{n,p} r^{\dM}$
for $r<1$.

The approximated volumes in Corollary~\ref{cor:vol} for the Stiefel manifold (with $p \neq n$)  and for the unitary group (with $p=n$) are compared with simulation in Fig.~\ref{fig:vol_un}. The figures are shown in logarithm scales since the region of interest is small chordal distance. We see that the approximations match almost exactly the simulations as $r\to 0$, which is a consequence of the exact volume normalization given in Theorem~\ref{thm_vol_stief}.

\subsection{Complementary Balls}
As an interlude, we discuss complementary balls. Small ball approximations can  be used to compute  the volume of very large  balls almost totally  covering the space.  
In the same manner as a sphere  can be fully covered by two complementary caps satisfying 
\begin{equation} \sigma(C_{D,R}(r)) =1 - \sigma(C_{D,R}(\sqrt{4R^2-r^2}) ) ,\end{equation}
we have the following analogous result for the Stiefel and Grassmann manifolds, proved in Appendix~\ref{proof:CompBall}. 
\begin{Lem}  \label{lem:large_ball_stief} 
In the Stiefel manifold $(\stief{C}{n}{p}, d_c)$, 
\begin{equation}
\mu(B(r))=1-\mu(B(\sqrt{4p-r^2}))
\end{equation}
implying that the volume is symmetrical at $\mu(B(\sqrt{2p}))=1/2$. 

For the Grassmann manifold $(\grass{C}{n}{p}, d_c)$,  
\begin{equation}
\mu(B(r))=1-\mu\left(B_{\bot}\left(\sqrt{\frac{2p(n-p)}{n} -r^2}\right)\right),
\end{equation}
where $B_{\bot}(\gamma) = \left\{ [\Pbf] \in \grass{C}{n}{p}  \; |
\quad  d_c([\Qbf],[\Pbf]) \leq \gamma   \right\}$ for an arbitrary
center $[\Qbf] \in \grass{C}{n}{n-p} $ and with chordal
distance\footnote{Note  
that here the definition of the chordal distance between two
Grassmannian planes of  
non-equal dimensions differs from~\cite{Dai08}, where  
one would have
$\mu(B(r))=1-\mu(B_{\bot}(\sqrt{p-r^2}))$, see~\cite{PWTC15}.} as
defined in~\eqref{eq:ChordDistGrassDetrProj}.

\end{Lem}

We remark that there is a structural difference here between the Stiefel and Grassmann manifolds. The antipodal on the embedding sphere to a point in the Stiefel manifold always belongs to the same Stiefel manifold. This is not always true for the Grassmann manifold except for $p=n/2$, as the antipodal on the sphere to a point in $\grass{C}{n}{p}$ belongs to $\grass{C}{n}{n-p}$.
This observation will be useful in the interpretation of the high-dimensional regime  discussed next. 

\begin{figure*}
\begin{center}
 \subfigure[Grassmann manifold $\grass{C}{n}{p}$  \label{fig:vol_grass_cap_approx}]{ \includegraphics[width=0.49\textwidth]{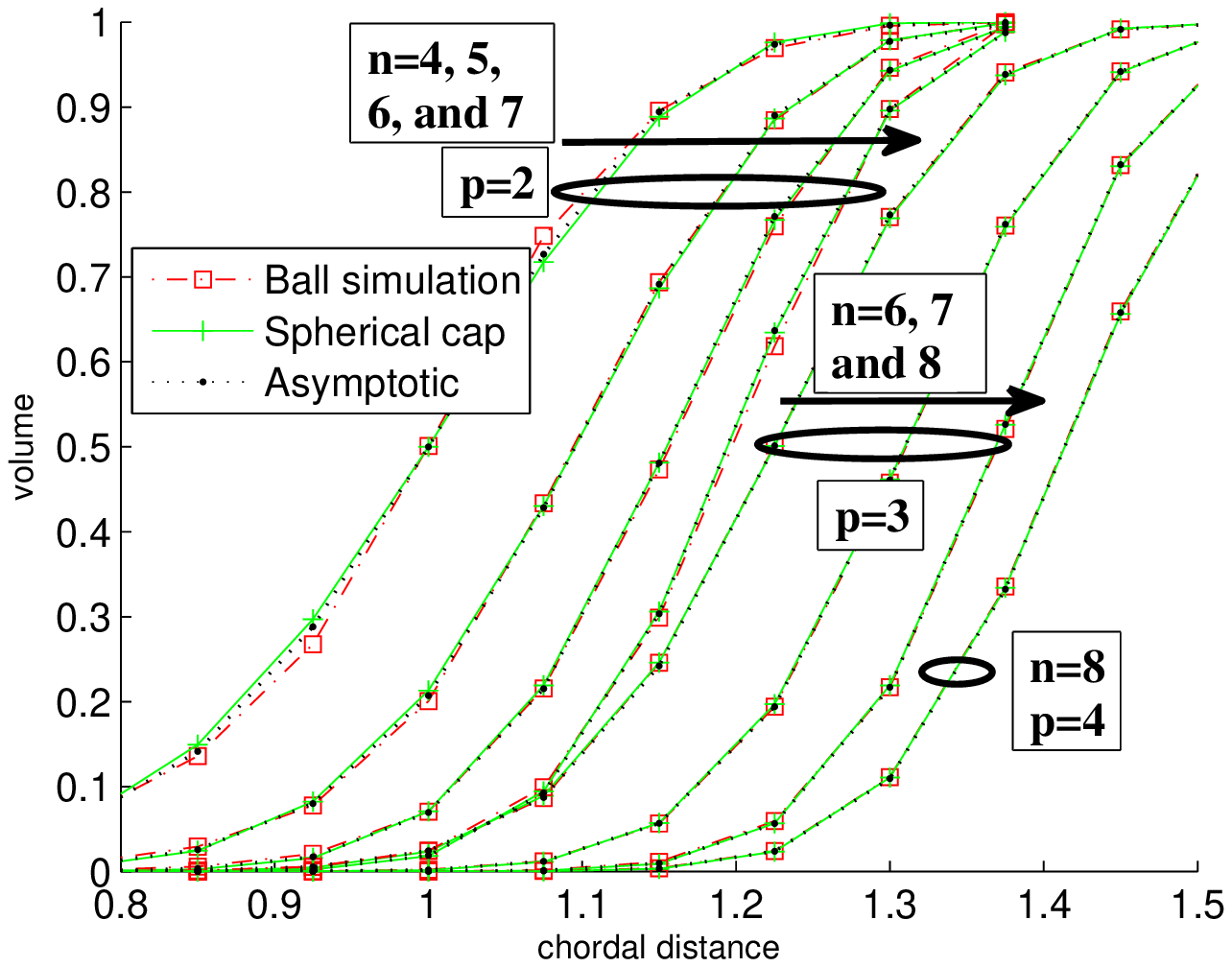}} \\
 \subfigure[Unitary group $\Ucal_n$]{\includegraphics[width=0.49\textwidth]{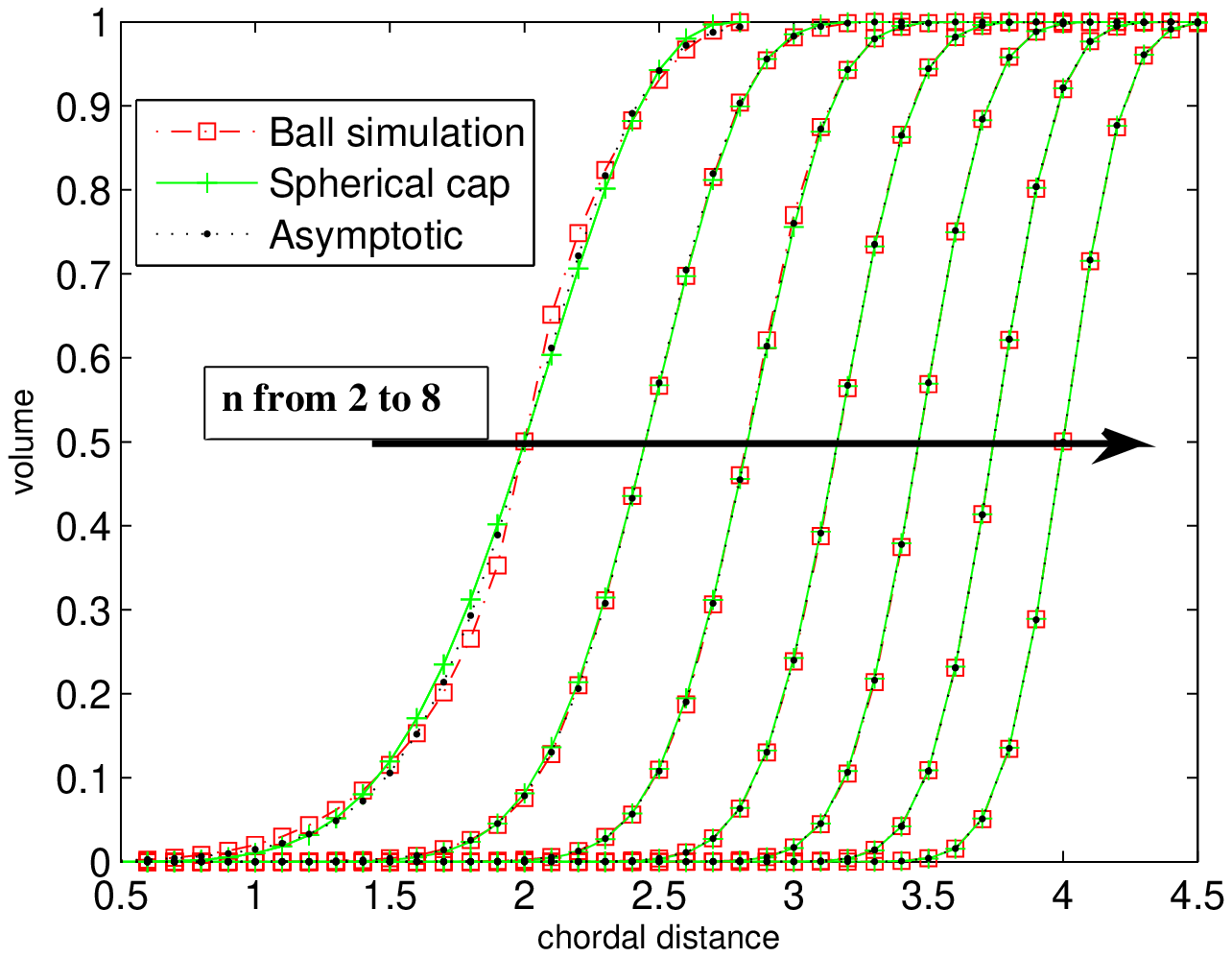}}
\subfigure[Stiefel manifold $\stief{C}{n}{p}$ ]{\includegraphics[width=0.49\textwidth]{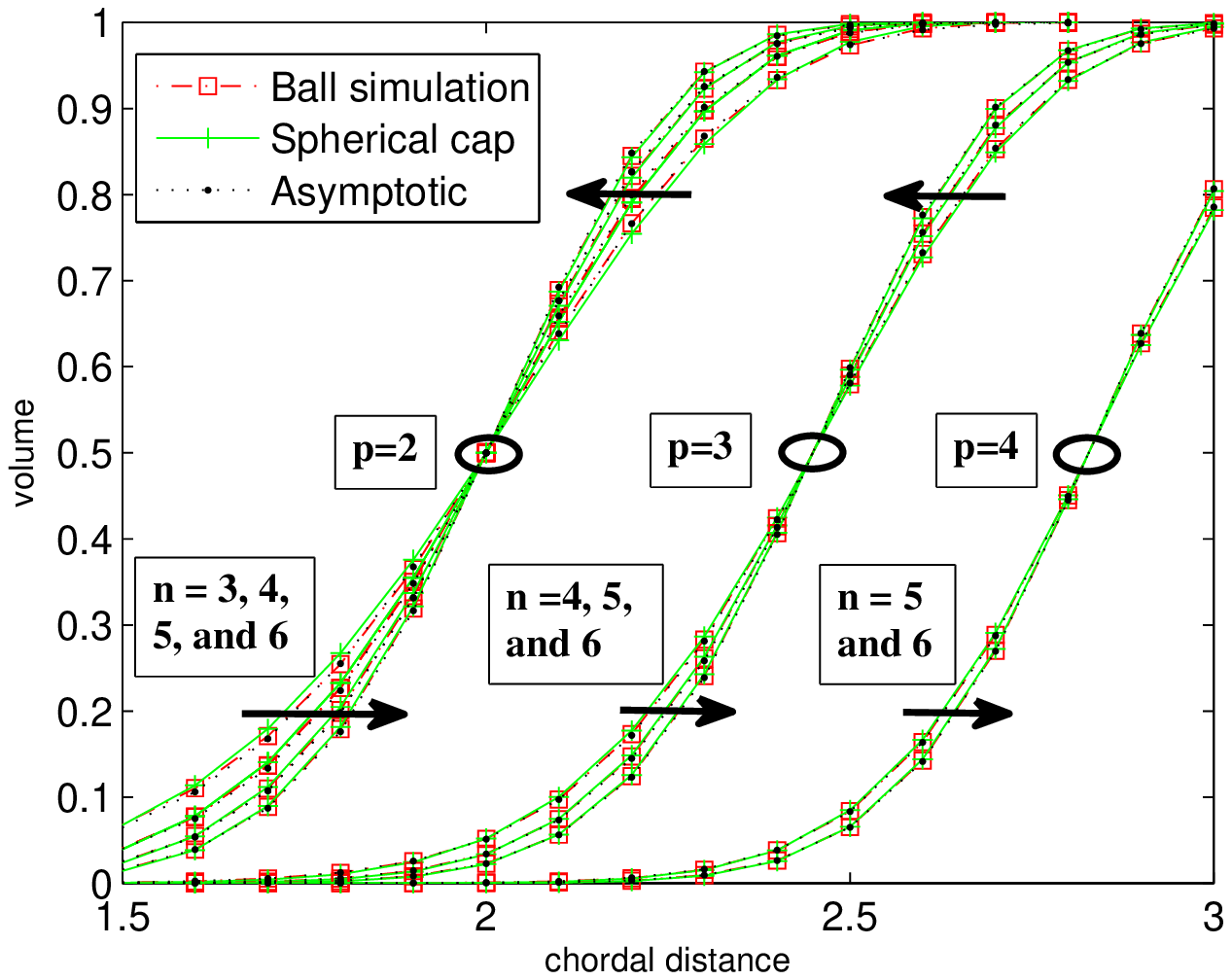}}
\end{center}
\vspace{-0.3cm}
  \caption{Illustration of Theorem~\ref{thm:cap_asympt}: volume of balls in manifolds (Monte Carlo simulation), spherical cap volume~\eqref{eq:vol_caps}, and asymptotic evaluation~\eqref{eq:largecap}  (equivalently~\eqref{eq:vol_asymp_stief1} and~\eqref{eq:vol_asymp_grass}).  
 \label{fig:vol_cap_approx}
}
\end{figure*}

\subsection{Volume Comparison Theorem and Spherical-Cap Approximation}
We now present a second volume approximation:   
For the manifolds $\Mcal$ isometrically embedded in $\Scal^D(R)$, the uniform measure of a ball in $\Mcal$  can be well approximated by the spherical measure of a cap on $\Scal^D(R)$. The two normalized volumes are indeed asymptotically equivalent in the high-dimensional regime. 
Before stating the theorem, we highlight an intermediate result  of independent interest. 
\begin{Lem} \label{lem:ChDistGaussian}
The squared chordal distance $d_c^2= d_c^2\left(\Ibf_{n,p},\Ybf\right)$ or $d_c^2= d_c^2([\Ibf_{n,p}],[\Ybf])$ drawn from a uniformly distributed random point
$\Ybf \in \stief{C}{n}{p}$ and a reference point, say $\Ibf_{n,p}$, converges in distribution to
a Gaussian random variable: 
\begin{equation} \frac{1}{ \sqrt{\var{d_c^2}}} \left( d_c^2-\expect{d_c^2}\right) \xrightarrow{d} \Ncal\left(0,1\right) ,  
\end{equation} 
\begin{itemize} 
\item  for  the Stiefel manifold $\stief{C}{n}{p}$, as $n\to \infty$  
and where the finite-size regime mean and variance are given by 
\begin{equation} 
\expect{d_c^2 }= 2p, \quad  \quad  \var{d_c^2} = \frac{2p}{n}; 
\end{equation}  

\item for  the Grassmann manifold $\grass{C}{n}{p}$,  as  $n,p \to \infty $ with  \mbox{$(n-2p)$} fixed, 
and where 
\begin{equation}
\expect{d_c^2} = \frac{p(n-p)}{n}, \quad \var{d_c^2} = \frac{p^2(n-p)^2 }{n^4-n^2 }.
\end{equation} 
\end{itemize} 
\end{Lem}

The proofs are given in Appendix~\ref{proof:ChDistGaussian}.  
They are obtained by reducing the chordal distances to the linear
statistics of random matrix ensembles and studying their
moment-generating functions. Such linear statistics are asymptotically
Gaussian, and by computing the two first moments one finds the
asymptotic forms. The case of the Stiefel manifold is closely related
to the partition function of the von-Mises Fisher distribution, while
for the Grassmann manifold it is related to the partition function of
the Bingham distribution~\cite{chikuse2012statistics}.

It follows that the chordal distance is well-approximated by a Gaussian random variable with the same finite-size mean and variance, i.e.  $d_c^2(\Ibf_{n,p},\Ybf) \sim  \Ncal\left(2p,\frac{2p}{n}\right)  $ for the Stiefel manifold $ \stief{C}{n}{p}$, and $d_c^2([\Ibf_{n,p}],[\Ybf]) \sim \Ncal\left(\frac{p(n-p)}{n}, \frac{p^2(n-p)^2 }{n^4-n^2 }\right)$ for the Grassmann manifold $\grass{C}{n}{p}$.

Using~\eqref{eq:largecap} for the hypespherical cap volumes, one gets a volume comparison of the ball in the embedded manifold with the embedding spherical cap:
\begin{Thm}
\label{thm:cap_asympt}
The normalized volume of a ball of radius $r$ in $\Mcal $  is
asymptotically equal  to the  
normalized
volume of a cap with same radius in the embedding sphere $(\Mcal,d_c) \hookrightarrow \Scal^{D-1}(R) $,
\begin{equation}
\mu(B(r)) \simeq  \sigma(C_{D,R}(r)) 
\end{equation}
in the high-dimensional regime as given in~\eqref{eq:largecap}. We have  
\begin{itemize} 
\item  for  the Stiefel manifold $\Mcal =\stief{C}{n}{p}$,
\begin{equation}
 \mu(B(r))   \simeq \frac12 \erf\left(\sqrt{np}\right) -\frac12  \erf\left(\sqrt{np}\left(1-\frac{r^2}{2p}\right)\right),
\label{eq:vol_asymp_stief1}
\end{equation}
as $n\to \infty$  with $\sqrt{2np}\left(1-\frac{r^2}{2p}\right)$ fixed;

\item  and for the Grassmann manifold $\Mcal =\grass{C}{n}{p}$, 
\begin{multline}
 \!\!\!\!\!\!\mu(B(r))   \simeq \frac12 \erf\left(\sqrt{\frac{n^2-1}{2}}\right) \\  -\frac12  \erf\left(\sqrt{\frac{n^2-1}{2}} \left(1-\frac{n \ r^2}{p(n-p)}\right)\right)
\label{eq:vol_asymp_grass}
\end{multline}
as $n,p \to \infty $ with $(n-2p)$  and $\sqrt{n^2-1} \left(1-\frac{n \ r^2}{p(n-p)}\right)$ fixed.
\end{itemize} 
\end{Thm}

The proof is given in Appendix~\ref{proof:VCTheorem}. The volume~\eqref{eq:vol_asymp_stief1}  generalizes our previous result~\cite{WPTC_ISIT15,WPCT15} from the unitary group to the Stiefel manifold. The volume~\eqref{eq:vol_asymp_grass} corresponds to the special case $p=q$ in~\cite{PWTC15}. 
Theorem~\ref{thm:cap_asympt} provides a geometric unification of the asymptotic expressions which are shown to be closely related due to their spherical embedding. 
The different asymptotic regimes for the Stiefel and Grassmann manifolds can be geometrically understood as the behaviors in Lemma~\ref{lem:large_ball_stief}. The Stiefel manifold fully covers all possible  distances on its embedding sphere. This is not the case for the Grassmann manifold, and as a consequence $\mu(B(r))$ and  $\sigma(C_{D,R}(r))$ are not defined on the same support, except for $\grass{C}{n}{\frac{n}{2}}$. In the regime $n,p \to \infty $ with fixed $n-2p$, one has $p \to \frac{n}{2}$ and the volume expressions are then asymptotically defined on the same support.  

There exists a body of literature on  comparison theorems in Riemannian geometry that compares volumes among manifolds with reference to the standard sphere. However, these theorems often compare manifolds of equal dimension, e.g. the Bishop-Gromov inequality or the Berger-Kazdan comparison theorem. 
To the best of our knowledge, volume comparisons of these manifolds with their embedding spheres in Theorem~\ref{thm:cap_asympt} is new. 

The quality of the approximation is illustrated in Figure~\ref{fig:vol_cap_approx}. The normalized volume of metric balls for different Grassmann and Stiefel manifolds obtained via Monte Carlo simulations is compared to the normalized volume of their respective embedding hyperspherical cap, together with their joint asymptotic expression.  The exact volume of the hyperspherical cap is as given by Lemma~\ref{Lem:vol_caps}.
This numerical evaluation shows that the three volume expressions are very close to each other even in the low dimensional regime.

\section{Kissing Radius and Density}

In this section, we discuss the kissing radius of codes with chordal
distance, and apply volume approximations derived in the previous
section to evaluate code density. Recall that for the Grassmann and
Stiefel manifold with chordal distance, the kissing radius $\varrho$
cannot be directly expressed in term of the minimum distance of the
code. In the following, upper and lower bounds on $\varrho$ and the
corresponding bounds on density are provided.

\subsection{Preliminaries}
\subsubsection{Hypothetical Covering Radius}
Ideally, a set of packing balls would fully cover the space, reaching
a maximum density of one.  
This is only possible when the cardinality of the code is $N=2$,
otherwise  
one gets an upper bound. This ideal radius $r_N$  
fulfills 
\begin{equation}
\label{eq:rn}
 \mu( B(r_N))=\frac{1}{N} .
 \end{equation} 

Two volume approximations were discussed in the previous section. 
Depending of the regime, one may compute $r_N$ accordingly.
\begin{itemize}
\item For $ N \geq c_{n,p}^{-1}$ the radius is less than one, and the small ball approximation leads to
\begin{equation}
\label{eq:rn_small_ball}
r_N \approx ( c_{n,p} N )^{\frac{-1}{\dim}},
 \end{equation} 
where $c_{n,p}$ is given in  
\eqref{eq:cnpS},~\eqref{eq:cnpG} 
and $\dim$ is the dimension of the manifold as given in  Table~\ref{tab:emb}.  For the Grassmann manifold, \eqref{eq:rn_small_ball} holds with equality with $ N \geq c_{n,p}^{-1}$~\cite{Dai08}. 

\item Otherwise for a larger ball, occurring with $n, p$ large and relatively small $N$, the spherical approximation of Theorem~\ref{thm:cap_asympt} leads to
\begin{equation}
\label{eq:rn_large_ball}
r_N \approx \sqrt{2} R \sqrt{\textstyle 1-\sqrt{\frac{2}{D}} \erf^{-1}\left(\erf\sqrt{\frac{D}{2}}-\frac{2}{N}\right)} ,
 \end{equation} 
where $R,D$ are the radius and the dimension of the spherical embedding, provided in  Table~\ref{tab:emb}. 

\end{itemize}

\subsubsection{Kissing Radius for Spherical Codes}
For spherical codes with chordal distance, the kissing radius of a code is given by a one-to-one mapping from the minimum distance, directly computable by the Pythagorean theorem.
Given an $(N,\delta)$-spherical code on  $\Scal^{D-1}(R)$, the midpoint on the geodesic between the two codewords of distance $\delta$ is at distance $\varrho_s$ from the extremities:
\begin{equation}
\label{eq:kiss_sph}
\varrho_s = \sqrt{2} R \sqrt{1-\sqrt{1-\frac{\delta^2}{4R^2}} },
\end{equation} 
which can be inverted as 
\begin{equation}
\label{eq:kiss_sph2}
\delta^2=4 \varrho_s^2-\frac{\varrho_s^4}{R^2}.
\end{equation}

\subsubsection{Preliminary Bounds on Kissing Radius}

Since we are considering a manifold isometrically embedded in the
Euclidean sphere $\Scal^{D-1}(R)$, an $(N,\delta)$-code in $\Mcal$ is 
an $(N,\delta)$-spherical code. As a consequence, as balls of radius
$\varrho_s$ are non-overlapping on $\Scal^{D-1}(R)$, their inverse
image on $\Mcal$ are also non-overlapping, and we can deduce that
$\varrho_s \leq \varrho$. On the other hand, we know that for every
non-overlapping ball of radius $r$, we have $r\leq r_N$. 
\begin{Lem}
\label{Lem:bd_kiss}
Given an $(N,\delta)$-code in $\Mcal$ isometrically embedded in $\Scal^{D-1}(R)$, the kissing radius $\varrho$  is bounded by  
\begin{equation}
 \varrho_s \leq \varrho \leq r_N,
\label{bd0:midpoint}
 \end{equation}
where  $\varrho_s$ is given in~\eqref{eq:kiss_sph} and $r_N$ satisfies~\eqref{eq:rn}.
\end{Lem}
It should be noted here that the lower bound is a function of the minimum distance $\delta$, while the upper bound is a function of the cardinality $N$.

\begin{figure*}
\begin{center}
 \subfigure[Grassmann manifold $\grass{C}{8}{4}$  \label{fig:KR_grass}]{ \includegraphics[width=0.49\textwidth]{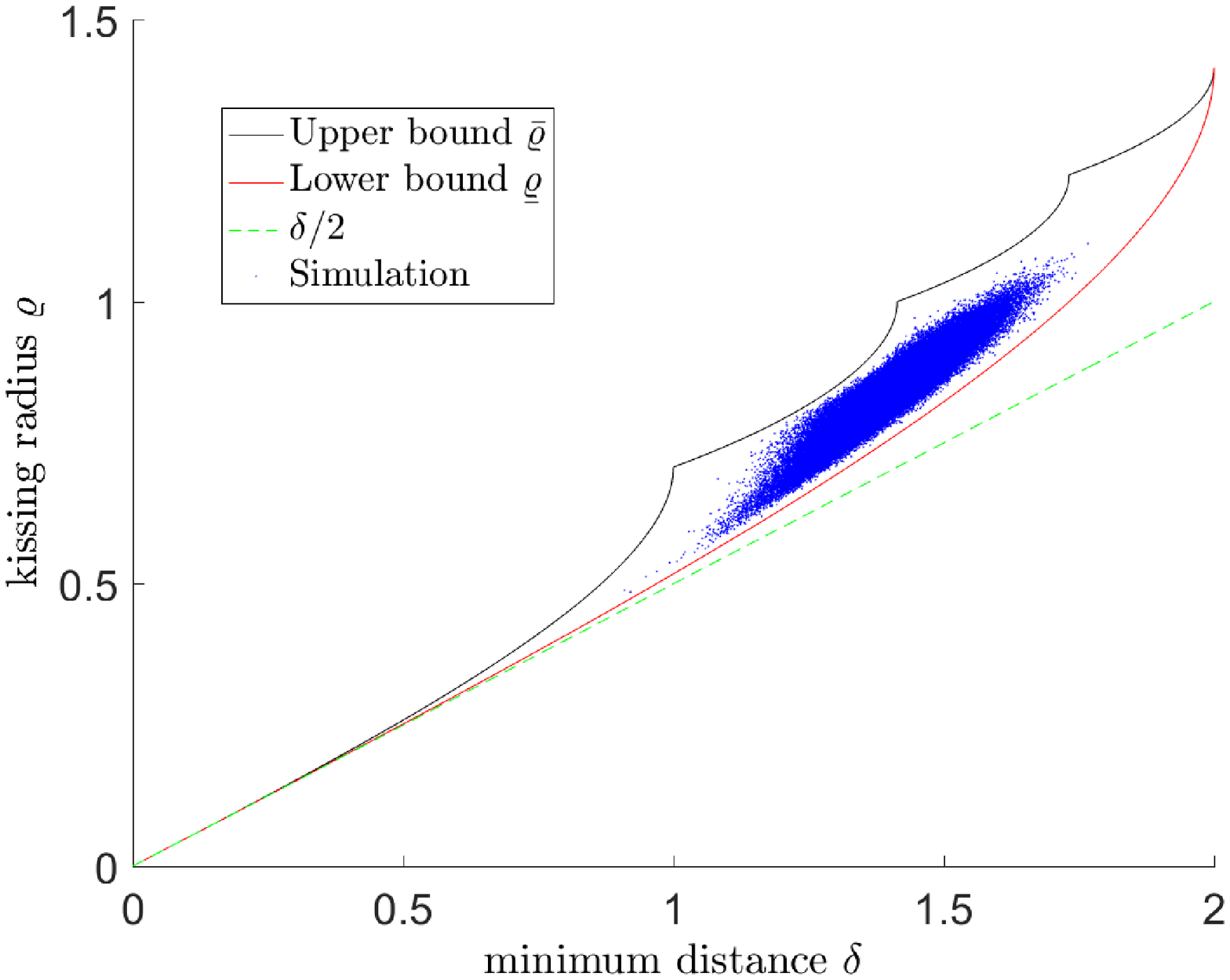}} \\
 \subfigure[Unitary group $\Ucal_3$]{\includegraphics[width=0.49\textwidth]{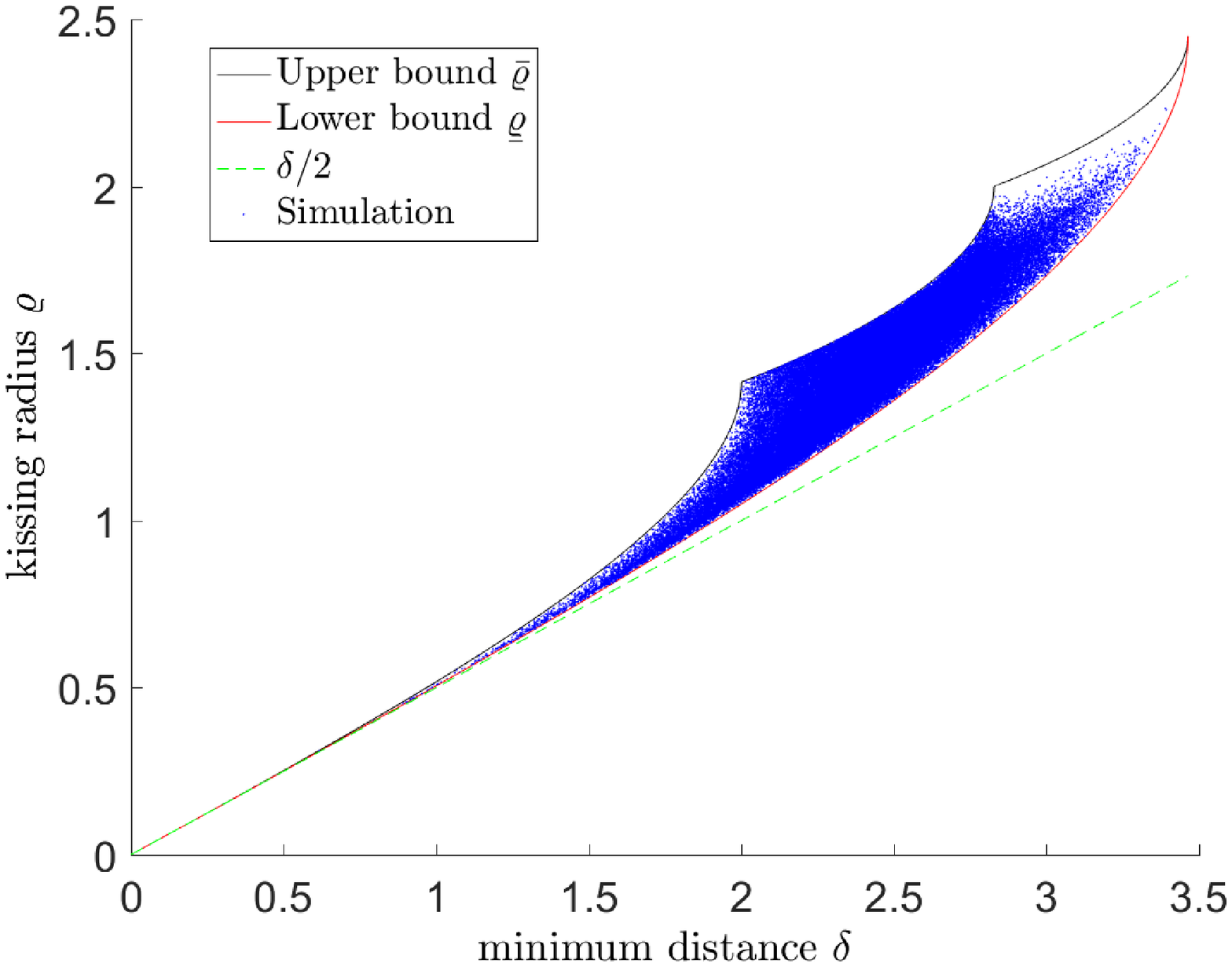}}
\subfigure[Stiefel manifold $\stief{C}{4}{2}$ ]{\includegraphics[width=0.49\textwidth]{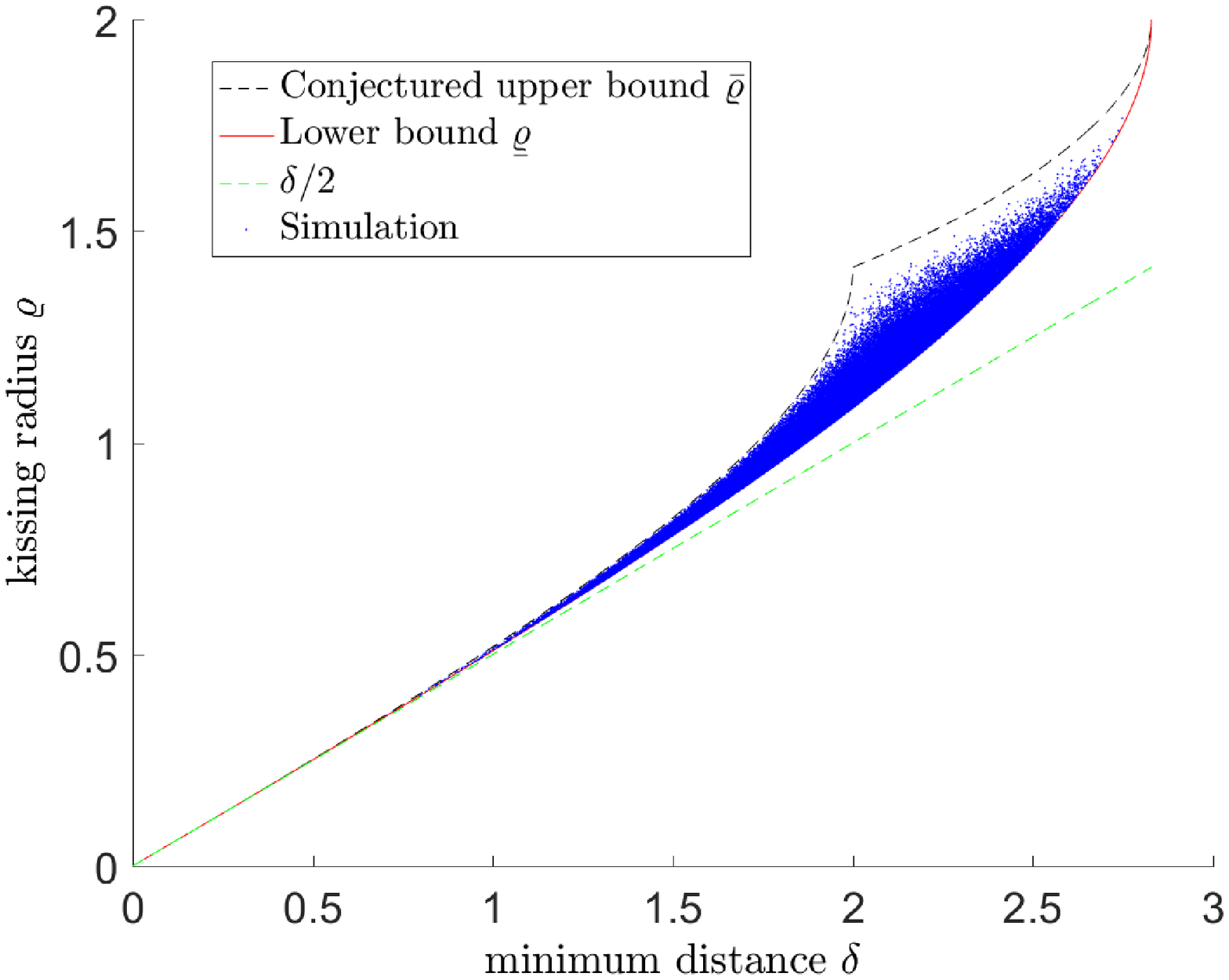}}
\end{center}
\vspace{-0.3cm}
  \caption{Illustration of the kissing radius bounds~\eqref{bd:midpoint} from  Proposition~\ref{prop:kissing_radius}. Bounds are compared to simulated midpoints between two randomly generated  codewords. It is also compared with the estimate $\delta/2$, corresponding to the classical packing radius in flat geometry.   
 \label{fig:midpointB}
}
\end{figure*}

\subsection{Bounds on Kissing Radius and Density}

We now provide bounds on the kissing radius as a function of the minimum distance of the code only, and corresponding bounds on code density.
\begin{Prop} \label{prop:kissing_radius}
For any   $(N,\delta)$-code $\Ccal$ in $\Mcal$, we have
\begin{equation} 
\label{bd:midpoint} 
  \rhol  \leq \varrho \leq   \rhou,
\end{equation}
where 
\begin{equation} 
\rhol  =  \left\{	\begin{array}{ll}  \sqrt{\frac{p}{2}\left(1- {\textstyle \sqrt{1-\frac{\delta^2}{p}}} \right) }  &  \text{ for } \Mcal = \grass{C}{n}{p} \\
									\sqrt{ 2 p  \left(1-\sqrt{1-\frac{\delta^2}{4p}} \right) }   &  \text{ for } \Mcal = \stief{C}{n}{p} \\
\end{array} \right. \label{eq:rhol}
\end{equation} 
\begin{equation} 
\rhou = \left\{	\begin{array}{ll}  \frac{1}{\sqrt{2}} \sqrt{ \lceil \delta^2 \rceil - \sqrt{ \lceil \delta^2 \rceil-\delta^2}  } &  \text{ for } \Mcal = \grass{C}{n}{p} \\
     	\sqrt{ 2} \sqrt{  \left\lceil \frac{\delta^2}{4} \right\rceil -\sqrt{  \left\lceil \frac{\delta^2}{4} \right\rceil - \frac{\delta^2}{4} }  } &  \text{ for }  \Mcal = \Ucal_{n}  
\end{array} \right. 
\label{eq:rhou}
\end{equation}
and $\lceil x \rceil$ is the smallest integer greater than $x$. It follows, therefore, that 
the density is bounded by 
\begin{equation} 
\label{eq:density}
N\mu \left( B(\rhol ) \right)  \leq \Delta(\Ccal) \leq \min \left \{ 1, N\mu \left( B(\rhou) \right) \right \} .
\end{equation}
\end{Prop}

A detailed proof can be found in Appendix~\ref{proof:kissing}. Given
two points either on the Grassmann manifold or the unitary group,
their midpoint can be determinated according to their principal
angles. The bounds then follow by optimizing over the principal
angles. For the unitary group, the obtained lower bound matches
the spherical lower bound in Lemma~\ref{Lem:bd_kiss}, and can thus be
extended to any Stiefel manifold. 
We conjecture that the upper bound~\eqref{eq:rhou} can be generalized to all Stiefel
manifolds. However, the lack of a closed-form geodesic equation between two points,
as explained in Remark~\ref{Rem:RiemLogMap} (or equivalently the
absence of the notion of principal angles), renders a tentative proof of
generalization difficult. Nevertheless, numerical
experiments support this generalization, and note that the upper
bounds~\eqref{eq:rhou} do not depend on any dimension parameter.

For the Grassmann manifold, the lower bound in Proposition~\ref{prop:kissing_radius} provides an improvement of the spherical embedding bound (see Appendix~\ref{Proof:ImprovedKRGrass}). 
\begin{Cor} \label{Corr:ImprovedKRGrass} For the Grassmann manifold, the lower bound in Proposition~\ref{prop:kissing_radius}  is tighter than the lower bound in Lemma~\ref{Lem:bd_kiss}. These two bounds are equal if and only if $p=n/2$. 
\end{Cor}

Figure~\ref{fig:midpointB} illustrates the upper and lower bounds~\eqref{bd:midpoint} on the kissing radius and shows that $\delta/2$ is a good approximation for $\varrho$ only when the minimum distance of the code is relatively small. In general, since the chordal distance is not strictly intrinsic, we have $\frac{\delta}{2} <  \varrho$. 
The bounds are also compared to $10^5$ simulated midpoints between a fixed center and uniformly-distributed random points. We stress that in this case the bounds are tight in the sense that it is always possible to construct two diagonal codewords fulfilling the bounds. 

\begin{figure*}[t]
\begin{center}
       \includegraphics[width=0.8\textwidth]{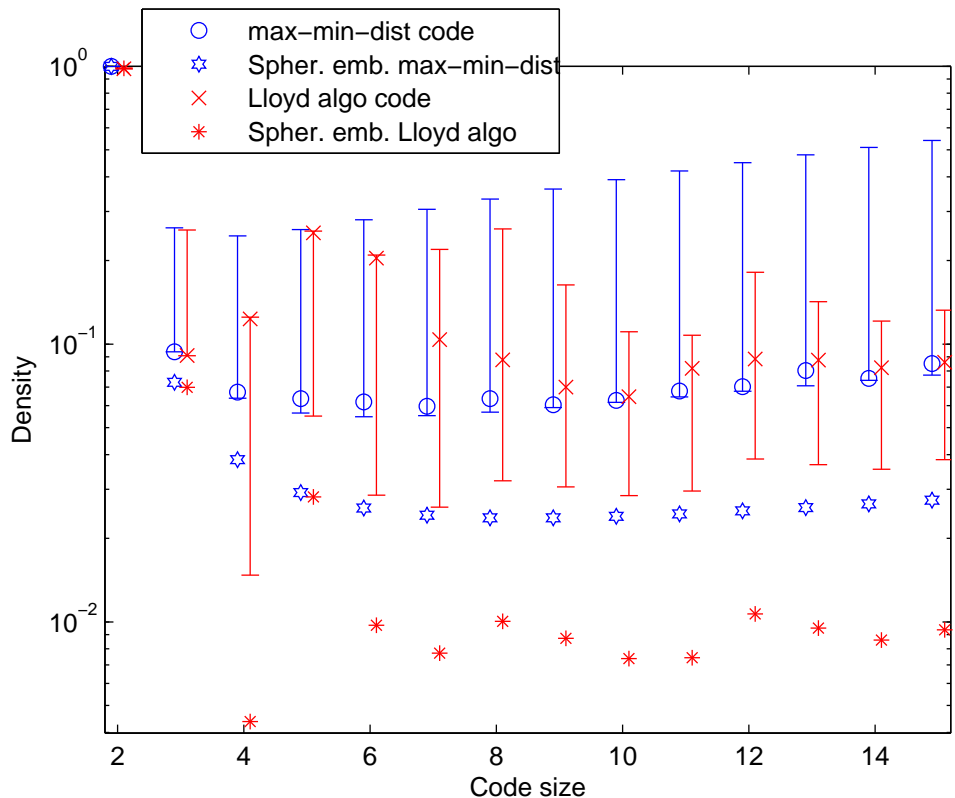}
\end{center}
\vspace{-0.5cm}
 \caption{Density of codes in $\grass{C}{4}{2}$ and their image code
   in $\Scal^{14}\left(\frac{1}{\sqrt{2}}\right)$ as a function of the number of
   points. The range bars represent the density bounds~\eqref{eq:density} from
   Proposition~\ref{prop:kissing_radius}   according to the minimum  distance of a code.
 \label{DensityG42Plot}
}
\end{figure*}

\subsection{On  High-Density Codes }
Since the lower bound~\eqref{eq:rhol} is an increasing function of  $\delta$, a large minimum distance always guarantees a good code density but not necessarily  the  highest one.  
\subsubsection{Numerical Experiments} 
Figure~\ref{DensityG42Plot} compares the density of codes with sizes
from $N=2$ to $N=15$ in $\grass{C}{4}{2}$ to the density bounds~\eqref{eq:density} in
Proposition~\ref{prop:kissing_radius}. The Grassmannian
$\grass{C}{4}{2}$ is embedded in $\Scal^{14}\left(\frac{1}{\sqrt{2}}\right)$,  
and the corresponding spherical Rankin bound\footnote{In this
  dimensionality, spherical codes achieve the simplex bound for $N\leq
  15$, then the orthoplex bound for $16\leq N\leq 30$.} 
provides an upper bound on the maximum possible minimum distance of
Grassmannian codes. We have generated Grassmannian codes according to
two criteria: 
a maximum-minimum-distance  
criterion, and a
low-distortion\footnote{The code distortion refers here to the average
  squared quantization error of a uniform random source quantized to
  the code.}  
criterion. For visibility, the densities
from the latter are displayed at $(N+0.1)$ while for the former at
$(N-0.1)$. 
For each code, two densities are displayed on Fig.~\ref{DensityG42Plot}: i) the density as a Grassmann code, and ii) the density as a spherical code. 
The bounds on Grassmann densities from Proposition~\ref{prop:kissing_radius} are
shown as range bars. The higher is the minimum distance of the code,
the higher is the bar.

Maximum-minimum-distance codes  
were obtained using the
Alternating Projection algorithm from~\cite{dhillon-2007}. The
obtained codes 
match the Rankin bound on squared minimum distance with $10^{-4}$
numerical precision. 
  
The Lloyd algorithm is used to create low-distortion codes.  
It is  expected to provide   
high density codes 
since the kissing radius corresponds to the first border effect of
Voronoi cells. In~\cite{Dai08}, it was shown that distortions
are bounded by idealized codes that would have only one border effect,
corresponding to a kissing radius equal to $r_N$, i.e. the upper bound~\eqref{bd0:midpoint} 
of Lemma~\ref{Lem:bd_kiss}.

While the Alternating Projection algorithm generates numerically
optimal simplicial codes with maximum minimum distance, their density
appears to be always close to the corresponding lower bound. This
means that the codes are strongly simplicial configurations where all
principal angles  
are almost equal. Comparing to the codes
generated by Lloyd algorithm, the results for cardinalities $3, 4$ and
$5$  
clearly show the weak relationship between maximizing the
minimum distance and maximizing the density. For $N=3$, the Lloyd
algorithm  
produced an equivalent max-min-distance code,  
which is at the
density lower bound. For $N=4$, it generated a code with  
smaller minimum distance but  
larger density, reaching   
the density 
upper bound. The
configurations obtained for $N=4$ correspond to codes numerically
equivalent to the   
closed form codes
in Table~\ref{tab:exampleCodeG42}.
For $N=5$, the Lloyd algorithm produced a code with minimum distance
close to the optimum, as for $N=3$, but this time with a density
reaching the upper bound, while the code from the Alternating
Projection algorithm is close to the lower bound. For other
cardinalities, the Lloyd algorithm produced codes with
greater or equal density but smaller minimum distance than
Alternating-Projection codes.

Lastly, an interesting   
characteristic 
of Grassmann codes is that their density for $N=2$ is not $\Delta=1$,
except for $p=\frac{n}{2}$. This comes from the fact that when $p \neq
\frac{n}{2}$ one cannot have two spherically antipodal points, c.f.
the symmetry of Lemma~\ref{lem:large_ball_stief}. For example, the
density of two orthogonal lines in $\grass{C}{3}{1}$ is equal to
$0.5$, and one can actually pack a third orthogonal line to reach the
density of $0.75$.

\subsubsection{Comments on Algebraic Constructions} 
For a given minimum distance $\delta$, the proof of
Proposition~\ref{prop:kissing_radius} shows that to achieve the largest
density in $\grass{C}{n}{p}$ or  $\Ucal_n$, a highly
concentrated distribution of principal angles should be targeted where
almost all non-zero principal angles should be at the maximum value
which is $\frac{\pi}{2}$  
for $\grass{C}{n}{p}$ and $\pi$ for $\Ucal_n$. Conversely, the lower bound is achieved when all principal angles are equal.

When constructing   
group codes from 
orbits of a symmetry group,   
the distribution of principal angles  
can  be controlled. 
The two infinite families of codes in
Table~\ref{tab:exampleCodesTensor} are  
orbit codes.  
$\Ccal^m_1$ is an orbit under the action of the 
$(m-1)$-fold tensor product of the projective unitary representation
of the Klein 4-group $V_4$~\cite{AsilomarOrbits}, i.e.  
a direct product
of the group of symmetries of a tetrahedron. As a result, the
principal angles between two points at minimum distance are  
$\left\{\arccos \frac{1}{\sqrt{3}},\arccos\frac{1}{\sqrt{3}}\right\}$ and  
the kissing radius is at the lower bound~\eqref{eq:rhol} of
Proposition~\ref{prop:kissing_radius}. The code $\Ccal^m_2$ is an
orbit of the fundamental representation of the symmetric group
$S_{2^m}$ via permutation matrices in the corresponding dimension. The
resulting principal angles are either $\left\{\frac{\pi}{2},
\frac{\pi}{2} \right\} $ or $\left\{0, \frac{\pi}{2}\right\} $ if the
two codewords have zero or one column in common, respectively.
Contrary to $\Ccal^m_1$, the codes $\Ccal^m_2$ reach the upper bound~\eqref{eq:rhou}
of Proposition~\ref{prop:kissing_radius}.

In~\cite{Calderbank99agroup-theoretic}, a family of optimal max-min
distance Grassmann codes is presented from which the code $\Ccal_2$ in
Example~\ref{Example1} is   
a subset of. These codes are orbits of a large Clifford group
generalizing the symmetries of the square. This results in highly
symmetric codes where many principal angles  
are $\frac{\pi}{2}$. Nevertheless, the kissing radius of these codes  
typically does not reach 
the upper bound~\eqref{eq:rhou}.  
Consider the   
half-subspaces construction in $\grass{C}{n}{\frac{n}{2}}$ from~\cite{shor1998family,Calderbank99agroup-theoretic}. 
Its complex extension~\cite{AshikhminCalderbankTIT10} leads to a code of
$2(n^2-1)$ codewords. Due to its orbit structure, the collection of
pairwise distances or principal angles can be computed independently
of the reference codeword as described
in~\cite{Calderbank99agroup-theoretic}. For each codeword, there is
always  
an antipodal codeword at pairwise distance $\delta^2_{i,j} = n$,
and $2(n^2-2)$ others at distance $\delta^2_{i,j} = \frac{n}{4}$.  
The codes thus meet the Rankin orthoplex bound. 
Among the codewords at minimum distance from a given codeword, there
are  $(n^2-4)$ that have 
principal angles $\{\frac{\pi}{2},\ldots, \frac{\pi}{2},0 ,
\ldots,0 \}$ (each value with multiplicity $\frac{n}{4}$), whereas the
remaining $n^2$ codewords have principal angles $\{\frac{\pi}{4},\ldots,
\frac{\pi}{4} \}$.  
Thus roughly half of the squared
mid-distances are  
$ \varrho^2_{i,j} = \frac{n}{4}$, and  
roughly the other half are  
$\varrho^2_{i,j} =\frac{(2-\sqrt{2})n}{8}$.  
These mid-distances thus meet
the upper bound~\eqref{eq:rhou} and lower-bound~\eqref{eq:rhol}, respectively. The kissing
radius of this code family is hence  
the lower bound 
$ \varrho=
\sqrt{\frac{(2-\sqrt{2})n}{8}}$. It should be remarked that a
mid-distance at the lower bound~\eqref{eq:rhol}, where principal angles between
two codewords are equal, corresponds  
to \emph{maximizing} the
diversity product for the given chordal distance. Therefore, about
half of the codeword pairs have a good product diversity of
$2^{-\frac{n}{2}}$, while the others have a product diversity  
$0$. The codes
in~\cite{Calderbank99agroup-theoretic,AshikhminCalderbankTIT10}
provide thus a mixture of high chordal mid-distance (as advocated
in~\cite{GoharyTIT09} by the chordal Frobenius-norm) and high product
diversity (as advocated in~\cite{HochwaldMarzettaTIT00}), in addition
to a high chordal distance (as advocated in~\cite{HanRosenthalGeo}).
This observation may provide further support to their good performance
observed in~\cite{AshikhminCalderbankTIT10} when applied to space-time
constellations.
 
\subsection{Relation to Density of Spherical Codes.}

As considered in this work, for manifolds isometrically embedded in Euclidean spheres, an $(N,\delta)$-code in $\Mcal$ is also an $(N,\delta)$-spherical code.  
The density of a code in the manifold differs from the density of its image code in the embedding sphere.  Combining the asymptotic equality from Theorem~\ref{thm:cap_asympt} between volume of balls in the manifold and hyperspherical cap with the kissing radius bounds from Proposition~\ref{prop:kissing_radius}, we have the following comparison.  
\begin{Prop} 
The density of codes in the Stiefel and Grassmann manifolds are asymptotically greater than the density of their image spherical codes in the high-dimensional regime of Theorem~\ref{thm:cap_asympt}.    
The inequality is strict for the Grassmann manifold with $N>2$. 
\end{Prop} 
This proposition is illustrated in Figure~\ref{DensityG42Plot} where densities of the image spherical codes of the considered Grassmann codes are also depicted. One observes that the densities of the image spherical codes are always less than the densities of the original Grassmannian codes for $N>2$. 

We remark also few special cases. In the case $p=1$, the lower and upper bounds on the kissing radius in Proposition~\ref{prop:kissing_radius} are matching $\rhol=\rhou$. 
Moreover with $p=1$, the Stiefel manifold is isometric to the embedding sphere, while for the Grassmann manifold the volume of a ball has been calculated exactly in~\cite{Dai08}.  In these cases the density can be computed exactly as a function of the minimum distance. 
\begin{Cor} For any packing with $p=1$, on manifolds  $\grass{C}{n}{1}$ or $\stief{C}{n}{1}$,
\begin{equation} 
\varrho= \rhol  =   \rhou 
\end{equation} 
and 
\begin{equation}
\label{eq:density_sphere}
\Delta(\Ccal)=\left\{	\begin{array}{ll}  N \left( \frac{1- \sqrt{1-\delta^2}}{2} \right)^{n-1} &  \text{ for } \Mcal = \grass{C}{n}{1} \\
			  N I_{\scriptscriptstyle \frac{1}{2} (1-\sqrt{1-\delta^2/4} )}\left(\frac{2n-1}{2}, \frac{2n-1}{2} \right) &  \text{ for } \Mcal = \stief{C}{n}{1}
\end{array} \right. .
\end{equation}
\end{Cor}
Finally, for the specific case of $n=2, \, p=1$, the Grassmann
manifold is also isometric to the real sphere: $\grass{C}{2}{1} \cong
S^2$~\cite[Ex.  17.23]{Bott}~\cite{TIT11}, and the discussed 
density of a code is consistent with the definition of the density for
a sphere packing  
from the literature~\cite{Clare}.

\subsection{Hamming-type Bounds}
Bounds on density directly translate to Hamming-type bounds on code cardinality and minimum distance.

\subsubsection{Hamming Bound on Cardinality}

According to Proposition~\ref{prop:kissing_radius}, we have:
\begin{Cor} For any $(N,\delta)$-code in $\Mcal$,  
and given $\rhol$ defined in \eqref{eq:rhol},
\begin{equation} 
\label{bd:hamNew}
N \leq \frac{1}{\mu(B(\rhol))} .
\end{equation} 
\end{Cor} 

For the unitary group, this bound is equivalent to the one derived in~\cite{Han}.  
For the Grassmann manifold,   
a bound based on  spherical embedding and asymptotic analysis was
provided in~\cite{Barg}. Equation~\eqref{bd:hamNew} provides a tighter bound.

\subsubsection{Hamming-type Bound on Minimum Distance}
From the spherical embedding of the manifolds,  we have the following Hamming-type bound on the minimum distance: 
\begin{Lem}
\label{eq:bd:dist}
Given an $(N,\delta)$-code in $\Mcal$ isometrically embedded in $\Scal^{D-1}(R)$, and a $r_N$ satisfying~\eqref{eq:rn}, we have 
\begin{equation}
\delta^2\leq 4 r_N^2-\frac{r_N^4}{R^2}.
\end{equation}
\end{Lem}
This follows from a direct combination of~\eqref{eq:kiss_sph2} and Lemma~\ref{Lem:bd_kiss}. 
For the unitary group, a similar bound was derived in~\cite[Theorem 2.4]{Han}. For the Stiefel manifold, the result is new. 
For the Grassmann manifold, a tighter bound is provided as a by-product of Proposition~\ref{prop:kissing_radius}, which is a generalization of a bound for line packing in~\cite{GianWelch} to any value of $p$: 
\begin{Lem} \label{dist_bound2} 
Given a $(N,\delta)$-code in $\grass{C}{n}{p}$,  
we have 
\begin{equation}  \label{bd:dist}
\delta^2 \leq 4 r_N^2  - \frac{4 }{p} r_N^4.
\end{equation} 
\end{Lem}

\begin{Rem} For the case $p=1$ the bound of Lemma~\ref{dist_bound2} reduces to the following bound derived in \cite[(32)]{GianWelch}:
\begin{equation}  
\delta^2 \leq 4 N^{\frac{-1}{n-1}}  - 4 N^{\frac{-2}{n-1}}.
\end{equation} 
\end{Rem}


\subsubsection{Illustration and Conjectured Improvement}

\begin{figure}[t]
\begin{center}
       \includegraphics[width=0.50\textwidth]{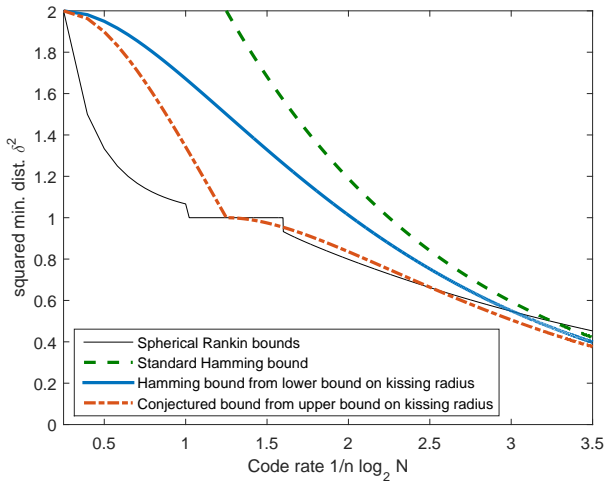}
\end{center}
\vspace{-0.3cm}
 \caption{The standard Hamming bound~\eqref{bd:hamStd}, the Hamming bound from the kissing radius analysis~\eqref{bd:dist} and a conjectured improvement~\eqref{eq:ConjBound}, compared to the Rankin bound for codes in $\grass{C}{4}{2}$.
 \label{fig:HBonMinDist}
}
\end{figure}

The standard Hamming bound~\eqref{bd:hamStd} and the Hamming bound from the kissing radius analysis~\eqref{bd:dist}  are displayed in Fig.~\ref{fig:HBonMinDist} for the Grassmannian $\grass{C}{4}{2}$ as a function of the code rate. One verifies that~\eqref{bd:dist} improves~\eqref{bd:hamStd} and they are getting equivalent as $N\to \infty$.  These Hamming bounds are also compared to the spherical Rankin bounds. 
Rankin provided three consecutive upper bounds in~\cite{Rankin} on the number of spherical caps that can be packed for a given angular radius. The two first bounds, known as the simplex and orthoplex bounds, can be easily inverted to  bound the minimum distance.  However, the third bound is expressed through an integral which we numerically inverted for Fig.~\ref{fig:HBonMinDist}. 
For $\grass{C}{4}{2}$, the Rankin bounds are tight at the saturating value $N=30$~\cite{PitavalICC11}, i.e. at rate $1.22$. For higher cardinality the Rankin bounds are not reachable.  As expected, the Hamming bounds become tighter than the Rankin bounds for large code sizes. 

By construction the Hamming bound is loose even if one would be able to exactly compute the kissing radius $\varrho$ as a function of minimum distance. 
The manifold can never be totally covered by packing balls (except for $N=2$), and one always has a gap between $\varrho$ and $r_N$. 
In order to sharpen the Hamming bound one could use $\rhou$ rather
than $\rhol$ to get a better estimate of the minimum distance. It is
very likely that in general $\rhou \leq r_N$, which can be verified at
the two ends of the code size spectrum $N=2$ and $N \to \infty$.
Inverting~\eqref{eq:rhou} gives the following approximation,  
\begin{equation}  
\delta^2 \lesssim    \left\{	\begin{array}{ll} \lceil 2 r_N^2 \rceil-\Big(\lceil 2 r_N^2 \rceil-2 r_N^2\Big)^2   &  \text{ for } \Mcal = \grass{C}{n}{p} \\ 
 4 \lceil  \frac{r_N^2}{2} \rceil-4\Big(\lceil \frac{r_N^2}{2} \rceil-\frac{r_N^2}{2}\Big)^2    &  \text{ for }  \Mcal = \stief{C}{n}{p}\end{array} \right. ,
\label{eq:ConjBound}
\end{equation} 
which is guaranteed to be an upper bound as $N \to \infty$. 
This conjectured improvement is displayed for $\grass{C}{4}{2}$ in Fig.~\ref{fig:HBonMinDist}. Here, $r_N$ is computed from~\eqref{eq:rn_small_ball}  from~\cite{Dai08} as in this case it is exact for any $N\geq 2 $. 
Interestingly, it meets the Rankin orthoplex bound at $N=32$, i.e. close to  $N=30$ where this bound is saturating.

\section{Conclusion}
The density of Grassmann and Stiefel codes with chordal distance has been investigated. The analysis pertains to treat the codes as subclasses of spherical codes since the chordal distance induces embeddings in Euclidean hyperspheres. The investigation is motivated by an exotic behavior: in this context maximizing the density of a code is not equivalent to maximizing its minimum distance. We addressed both  critical steps to compute a code's density, which are computing the volume of balls and the kissing radius. For the volume of balls, our main results included a proper scaling of the manifold volume consistent with the chordal distance, which is needed in small ball approximation. Moreover, an asymptotic Gaussian behavior of chordal distances in the high-dimensional regime leading to an asymptotic equivalence between the volume of balls in the manifolds and the volume of caps in the embedding spheres was found. Then the kissing radius and density of codes were bounded as a function of the minimum distance of the codes. It was concluded that Stiefel and Grassmann codes have larger density than their image spherical codes in high dimensions.  Finally, as a by-product of the analysis, refinements of the standard Hamming bounds for Stiefel and Grassmann codes were provided.

\appendices

\section{Proof of Lemma~\ref{Lem:vol_caps}--Hyperspherical Cap Area}
\label{proofs:Cap}
\subsection{Proof of~\eqref{eq:vol_caps}}

The area of a hyperspherical cap is given in~\cite{li_spher_cap} in which the radius of the cap is measured by an angle~$0\leq \phi \leq \frac{\pi}{2}$. This translates in our notation to a radius  of $r^2 = 2 R^2 (1-\cos \phi )$, that gives $\sin^2 \phi =\frac{r^2}{R^2} (1-\frac{r^2}{4R^2})  $ and  directly leads to 
\begin{equation}
\label{eq:vol_capsAp}
 {\rm vol }(C_{D,R} (r) )  = \frac{A_D(R)}{2} I_{\frac{r^2}{R^2} (1-\frac{r^2}{4R^2})}\left(\frac{D-1}{2}, \frac{1}{2}\right) 
\end{equation}
for $r \leq \sqrt{2}R$. 

The result further simplifies by normalizing by the overall area of the sphere $A_D(R)$ and using the identity $I_{x}(a, a) = \frac12 I_{4x(1-x)}\left(a, \frac{1}{2}\right) $ valid for $0\leq x\leq \frac12$~\cite[(8.17.6)]{NIST:DLMF}\cite{Olver:2010:NHMF}. The volume for larger radius $r \geq \sqrt{2}R$ can be computed by the complementary cap, i.e. 
\begin{equation}\sigma(C_{D,R}(r) ) = 1-\sigma(C_{D,R}( \sqrt{4R^2 -r}) )  ,\end{equation}
and using the identity
$I_{x}(a, b) = 1- I_{1-x}(b,a) $~\cite[(8.17.4)]{NIST:DLMF}\cite{Olver:2010:NHMF}. Altogether, one obtains~\eqref{eq:vol_caps} for the whole range of the radius $0 \leq  r \leq 2R$.

\subsection{Proof of~\eqref{eq:smallcap} -- Small Cap}
We start from~\eqref{eq:vol_capsAp} which after normalization gives
\begin{equation} \sigma(C_{D,R}(r)) = \frac{1}{2} I_{\frac{r^2}{R^2} (1-\frac{r^2}{4R^2})}\left(\frac{D-1}{2}, \frac{1}{2}\right). 
\end{equation}
Expressing the regularized incomplete beta function  $I_{x}(a, b) = B_{x}(a, b)/ B(a, b) $
with the Beta function $B(a, b) = \Gamma(a) \Gamma(b) /\Gamma(a+b)$ and the incomplete Beta function  $B_{x}(a, b)$ gives
\begin{equation} \sigma(C_{D,R}(r)) = \frac{1}{2 \sqrt{\pi} } \frac{\Gamma(\frac{D}{2})}{\Gamma(\frac{D-1}{2}) }  B_{\frac{r^2}{R^2} (1-\frac{r^2}{4R^2})}\left(\frac{D-1}{2}, \frac{1}{2}\right).
\end{equation}
From the identity $B_{x}(a, b)=\frac{x^a}{a} F(a,1-b;a+1;x)$~\cite[(8.17.7)]{NIST:DLMF} where $F⁡(a,b;c;z)$ is the hypergeometric function given by~\cite[(15.2.1)]{NIST:DLMF}, it can be verified that  $B_{x}(a, b) =\frac{x^a}{a} (1+ O(x) ) $ as $x \to 0$, which in turn gives 
\begin{equation} 
\sigma(C_{D,R}(r)) = \frac{1}{2\sqrt{\pi}} \frac{\Gamma(\frac{D}{2})}{\Gamma(\frac{D+1}{2}) }   \frac{r^{D-1}}{R^{D-1}} (1+ O(r^2) )
\end{equation}
as $r \to 0$.

\subsection{Proof of~\eqref{eq:largecap} -- Large Dimension}
The normalized volume of a cap $C_{D,R}(r)$  can be interpreted as the probability that a uniformly distributed vector $\xbf \in \Scal^{D-1}(R)$ is at distance less than $r$ from a fixed point $\ebf$.   
Without loss of generality, we choose the center of the cap to be $\ebf= [R , 0,\ldots,0]^T$  and express the chordal distance by $d^2_c(\ebf, \xbf)=2R^2\left(1-x_{1,D}/\sqrt{D} \right)$ where $x_{1,D}= \frac{\sqrt{D}}{R^2}\langle \ebf, \xbf \rangle $ is the first coordinate of $\xbf$ normalized by the sphere radius and scaled. Noting that  
$$
0 \leq d_c(\ebf, \xbf) \leq r ~~\Leftrightarrow~~ 
\sqrt{D}  \geq x_{1,D} \geq \ \sqrt{D} \left( 1-\frac{r^2}{2 R^2}\right)~, 
$$ 
the volume is 
\begin{eqnarray}
\!\!\!\!\!\! \sigma(C_{D,R}(r)) \!\!&\!\!=\!\!&\!\! {\rm Pr}\{ \xbf\in \Scal^{D-1}(R) \, | \, 0 \leq d_c(\ebf, \xbf) \leq r \} \nonumber \\ 
									 \!\!&\!\!=\!\!&\!\! F_{x_{1,D}}\!\! \left(\sqrt{D} \right) -  F_{x_{1,D}}\!\! \left(\!\! \sqrt{D} \left(1-\frac{r^2}{2R^2}\right)\! \! \right) \label{eq:VolCapCDF}
\end{eqnarray}
where $F_{x_{1,D}}$ is the cumulative distribution function of the random variable $x_{1,D} $. Note that since the constraint $ 0 \leq d_c(\ebf, \xbf) $  is always satisfied, one has $F_{x_{1,D}}\left(\sqrt{D} \right) =1$. 

With $D\to \infty$, $x_{1,D}$ converges in distribution to a standard Gaussian random variable  $\sim \Ncal\left(0,1\right) $~\cite{Borel,Richards}, i.e. 
\begin{equation}
\lim_{D\to \infty }F_{x_{1,D} }(z)  =  \frac{1}{2} \left(1+ \erf\left( \frac{z}{\sqrt{2}} \right) \right) \\
\end{equation}
where 
\begin{equation}
\label{eq:erf}
\erf(x)=\frac{2}{\sqrt{\pi}}\int_{0}^{x}e^{-t^{2}} dt
\end{equation}
is the Gauss error function.  
Therefore, with $z=\sqrt{D} \left(1-\frac{r^2}{2R^2}\right)$ fixed, we have 
\begin{eqnarray} 
\!\!\!\!\! \lim_{D\to \infty }\sigma(C_{D,R}(r)) \!\!&=&\!\!  1 - \lim_{D\to \infty } F_{x_{1,D}}\left(z\right) \nonumber \\
                                      \!\!&=&\!\!  \frac{1}{2}- \frac{1}{2} \erf\left(\!\!  \sqrt{\frac{D}{2}} \left(1-\frac{r^2}{2R^2}\right)\!\!   \right) \! .
\end{eqnarray} 

Finally, in order to provide a finite-size approximation that satisfies the basic property of a measure, $\sigma(C_{D,R}(0))=0$, in addition to converging to the asymptotic form above, we use in~\eqref{eq:VolCapCDF} the finite-size correction $F_{x_{1,D}}(\sqrt{D}  ) \approx  \frac{1}{2} \left(1+ \erf\left( \sqrt{D/2}  \right) \right) \to 1 $ which corresponds to approximating $x_{1,D}$ to be Gaussian also in the finite-size regime. This leads to the given expression.

\section{Proofs of Overall Volumes~\label{proof:overall_volume}}

$\Mcal$ is an $m$-dimensional Riemann manifold with infinitesimal
metric  
\begin{equation}
ds^2 = \sum_{j,k=1}^m g_{jk} dx_j dx_k
\end{equation}
in local coordinates $\{x_j \}^m_{j=1}$. The volume element of $\Mcal$ is~\cite{pastur}
\begin{equation}
d\omega = \sqrt{\det \{ g_{jk} \}^m_{j,k=1}} dx_1 \ldots dx_m.
\end{equation}

Now, consider the Euclidean space of $m \times m$ complex matrices with its canonical inner product $(\mathbb{C}^{m \times m}, \langle \cdot,\cdot\rangle= \mathfrak{R}\Tr \cdot^H \cdot )$ from which the Riemann metrics considered are constructed.  Given $\Mbf \in \mathbb{C}^{m \times m}$ we have
\begin{eqnarray}
ds^2 &=&\mathfrak{R}\Tr( d\Mbf^H d\Mbf) = \| d\Mbf\|^2_F \nonumber \\
 &=& \sum^m_{j,k=1} \mathfrak{R}(d\Mbf_{jk})^2+\mathfrak{I}(d\Mbf_{jk})^2.
\end{eqnarray}
Restricting the metric to the space of skew-Hermitian  
matrices, with $\Abf \in \mathfrak{u}(m)$ we get
\begin{equation}
ds_{\Abf}^2 = - \Tr( d\Abf^2 )= \sum^m_{j=1} |d\Abf_{jj}|^2 +2\sum_{j <k} |d\Abf_{jk}|^2
\end{equation}
and the corresponding volume element
\begin{equation}
d\omega_{\Abf} = 2^{\frac{m(m-1)}{2}} \prod^m_{j=1} |d\Abf_{jj}| \prod_{j <k}^m  \mathfrak{R}(d\Abf_{jk}) \mathfrak{I}(d\Abf_{jk}).
\end{equation}
Note  here that the off-diagonal elements are counted twice, leading to an overall scaling factor of $2^{\frac{m(m-1)}{2}}$. 
The overall volumes of the manifolds need to be scaled accordingly to be consistent with the chosen metric. 

\subsection{Volume of Unitary Group}
Given a unitary matrix $\Ubf\in \Ucal_n$, by differentiating $\Ubf^H \Ubf = \Ibf$, we obtain
\begin{equation}
\Ubf^H d\Ubf +d\Ubf^H \Ubf = 0
\end{equation}
showing  that the differential form $\Ubf^H d\Ubf$ is skew-Hermitian. 
Due to the unitary invariance of the metric $ds^2$, its restriction to $\Ucal_n$ can be  expressed in terms of the global form $ \Ubf^H d\Ubf$. 
Then, the infinitesimal metric is
\begin{eqnarray}
ds_{\Ubf}^2 &=& - \Tr(\Ubf^H d\Ubf)^2  \nonumber \\
 &=&\sum^n_{j=1} |(\Ubf^H d\Ubf)_{jj}|^2 +2\sum_{j <k} |(\Ubf^H d\Ubf)_{jk}|^2
\end{eqnarray}
and the volume form in local coordinates is
\begin{equation}
d\omega_{\Ubf} = 2^{\frac{n(n-1)}{2}} d\nu_{\Ubf},
\end{equation}
where 
\begin{equation}
d\nu_{\Ubf} = \prod^n_{j=1} |(\Ubf^H d\Ubf)_{jj}| \prod_{j <k}^n  \mathfrak{R}((\Ubf^H d\Ubf)_{jk}) \mathfrak{I}((\Ubf^H d\Ubf)_{jk})
\end{equation}
is a common volume element normalized so that $\int_{\Ucal_n}  d\nu_{\Ubf}  = \frac{2^n \pi^{n^2}}{\widetilde{\Gamma}_n(n)}$~\cite{Muirhead,EdelmanPHD}. Finally with the metric considered, 
\begin{equation}
 \text{vol }\Ucal_n  =   \frac{2^{\frac{n(n+1)}{2}} \pi^{n^2}}{\widetilde{\Gamma}_n(n)},
\end{equation}
where the complex multivariate gamma function is
\begin{equation} \widetilde{\Gamma}_p(n)= \pi^{\frac{p(p-1)}{2}}
\prod_{i=1}^p \Gamma(n-i+1). \label{eq:MultiGammaC}
\end{equation}

\subsection{Volume for Stiefel Manifold in Theorem~\ref{thm_vol_stief}}

Now let $\Ybf \in \stief{C}{n}{p}$ and $\Ubf \in \Ucal_n$ such that $\Ubf^H \Ybf = \Ibf_{n,p}$, i.e. the first $p$ columns of $\Ubf =\begin{pmatrix} \Ybf & \Ybf^{\bot}  \end{pmatrix}$ are the columns of $\Ybf$. The differential form $\Ubf^H d\Ybf$ is ``rectangular skew-Hermitian'', i.e. $\Ubf^H d\Ybf =  \begin{pmatrix} \Ybf^H d\Ybf \\ \Ybf^{\bot H} d\Ybf \end{pmatrix}$ where $\Ybf^H d\Ybf $ is $p$-by-$p$ skew-Hermitian.
Similarly, due to unitary invariance of the metric, the volume element for the Stiefel manifold can be  expressed in terms of the global form $ \Ubf^H d\Ybf$ and is given in local coordinates as
\begin{equation}
d\omega_{\Ybf} = 2^{\frac{p(p-1)}{2}} d\nu_{\Ybf},
\end{equation}
where 
\begin{multline}
d\nu_{\Ybf}  = \prod^p_{i=1} |(\Ubf^H d\Ybf)_{ii}|  \\
\times \prod_{k=1}^p \prod_{j = k+1}^n \! \! \! \mathfrak{R}((\Ubf^H d\Ybf)_{jk}) \mathfrak{I}((\Ubf^H d\Ybf)_{jk})
\end{multline}
is a common volume element normalized so that $\int_{\stief{C}{n}{p}}  d\nu_{\Ybf}  = \frac{2^p \pi^{np}}{\widetilde{\Gamma}_p(n)}$~\cite{Muirhead,EdelmanPHD}. Finally, 
\begin{equation}
 \text{vol }\stief{C}{n}{p}  =   \frac{2^{\frac{p(p+1)}{2}} \pi^{np}}{\widetilde{\Gamma}_p(n)}.
\end{equation}

\subsection{Volume for Grassmann Manifold}
The volume of the Grassmann manifold directly follows from the quotient geometry over the Stiefel manifold associated with $d_g$: \begin{equation}\text{vol }  \grass{C}{n}{p} = \frac{\text{vol }  \stief{C}{n}{p}}{\text{vol }  \Ucal_p}. \end{equation} 

Alternatively it can be computed from the quotient geometry over the unitary group associated with $d_{g*}$
\begin{equation} \text{vol }  \grass{C}{n}{p} = 2^{-\frac{\dim}{2}} \frac{\text{vol }  \Ucal_n}{\text{vol }  \Ucal_p \text{vol }  \Ucal_{n-p}}, \end{equation} 
where the scaling coefficient comes from the $\sqrt{2}$ in the definition of $d_{g*}$ compared to $d_g$. 


\section{Proof of Lemma~\ref{lem:large_ball_stief}--Complementary Balls~\label{proof:CompBall}}

\paragraph{Stiefel Manifold} Consider the center to be $\Ibf_{n,p} \in
\stief{C}{n}{p}$. It has a unique antipodal point (i.e. a 
farthest possible point 
from $\Ibf_{n,p}$) which is $-\Ibf_{n,p}$
and we have
\begin{equation}
 d_c(\Ibf_{n,p},-\Ibf_{n,p}) =2\sqrt{p}  \triangleq d_{\max}. 
\end{equation}
Given a point $\Ybf\in \stief{C}{n}{p}$ such that $d_c(\Ibf,\Ybf) \geq r$, it is a direct verification that $d_c(-\Ibf,\Ybf) \leq \sqrt{d_{\max}^2-r^2}$, thus $\Ybf \notin B_{\Ibf}(r)$ implies  $\Ybf \in B_{-\Ibf}(\sqrt{d_{\max}^2-r^2})$ and finally 
\begin{equation} \mu(B(r))+ \mu(B(\sqrt{d_{\max}^2-r^2}))=1.
\end{equation}

\paragraph{Grassann Manifold} 
From the mapping~\eqref{eq:DetracedProj}, the Grasmmann manifolds $\grass{C}{n}{p}$ and $\grass{C}{n}{n-p}$  are both embedded in the same sphere $\Scal^{n^2-2}({ \scriptscriptstyle \sqrt{ \frac{p(n-p)}{2n}}})$.  
The definition of chordal distance thus directly extends between  any $[\Ybf]\in \grass{C}{n}{p} $  and $[\Zbf] \in \grass{C}{n}{n-p}$ as 
\begin{eqnarray}
d^2_c([\Ybf], [\Zbf])&=& \|\bar{\Pi}_{\Ybf}- \bar{\Pi}_{\Zbf}\|_F^2 \nonumber\\ 
 &=& \frac{2p(n-p)}{n} -\| \Ybf^H\Zbf \|_F^2.
\end{eqnarray}  
The distance reaches its maximum at $d_{\max} = \sqrt{\frac{2p(n-p)}{n}}$ with $[\Zbf ] = [\Ybf_{\perp}]$, where the $ \Ybf_{\perp}$ is the orthogonal complement of $ \Ybf$ such that $\begin{pmatrix} \Ybf & \Ybf_{\perp}   \end{pmatrix} $ is a unitary matrix. The maximum distance is exactly twice the embedding radius, $d_{\max} = 2 R$, i.e.  $ [\Ybf_{\perp}]$ is the antipodal of $[\Ybf]$ on the embedding sphere. 
The final result follows by the same argumentation as for the Stiefel manifold. Here the  Pythagorean theorem gives  $d^2_c([\Ybf], [\Zbf]) + d^2_c([\Ybf_{\perp}], [\Zbf]) = \frac{2p(n-p)}{n} $.

\section{Proof of the asymptotic Gaussianity of the chordal distances in Lemma~\ref{lem:ChDistGaussian}}
\label{proof:ChDistGaussian}

Before proceeding the proof, we first state some useful definitions and intermediary results.

\subsection{The Hypergeometric Function of Complex Matrix Argument}

For an $n\times n$ Hermitian matrix $\Abf$, the hypergeometric function of a complex matrix argument is defined as~\cite{1964James,1997Mathai}
\begin{multline}\label{eq:MHF}
_{p}\widetilde{F}_{q}\left(a_{1},\dots,a_{p};b_{1},\dots,b_{q};\Abf\right)= \\
\sum_{k=0}^{\infty}\sum_{\kappa}\frac{\left(a_{1}\right)_{\kappa}\cdots\left(a_{p}\right)_{\kappa}}{\left(b_{1}\right)_{\kappa}\cdots\left(b_{q}\right)_{\kappa}}\frac{C_{\kappa}(\Abf)}{k!},
\end{multline}
where $\kappa$ denotes a partition of integer $k$ into no more than
$n$ parts, i.e. $k=\kappa_{1}+\kappa_{2}+\dots+\kappa_{n}$ with
$\kappa_{1}\geq\kappa_{2}\cdots\geq\kappa_{n}\geq0$,
the sum is over all partitions,
and
\begin{equation}\label{eq:MHC}
(a)_{\kappa}=\prod_{j=1}^{n}(a-j+1)_{\kappa_{j}}=\prod_{j=1}^{n}\frac{\left(\kappa_{j}+a-j\right)!}{\left(a-j\right)!}
\end{equation}
is the multivariate hypergeometric coefficient~\cite[Eq.~(84)]{1964James}. In~(\ref{eq:MHF}), $C_{\kappa}(\Abf)$ denotes a zonal polynomial~\cite{1964James,1997Mathai}, which is a homogenous symmetric polynomial of degree $k$ in the $n$ eigenvalues of $\Abf$. Denoting  the $j$-th eigenvalue of $\Abf$ by $a_{j}$, the zonal polynomial can be represented as~\cite[Eq.~(85)]{1964James},
\begin{equation}
C_{\kappa}(\Abf)=\chi_{\kappa}(1)\chi_{\kappa}(\Abf),
\end{equation}
where
\begin{equation}
\chi_{\kappa}(1)=\frac{k!\prod_{1\leq i<j\leq n}(\kappa_{i}-\kappa_{j}-i+j)}{\prod_{j=1}^{n}(\kappa_{j}+n-j)!}
\end{equation}
and
\begin{equation}\label{eq:Schur}
\chi_{\kappa}(\Abf)=\frac{\det\left(a_{i}^{\kappa_{j}+n-j}\right)}{\det\left(a_{i}^{n-j}\right)}
\end{equation} 
is a Schur polynomial\footnote{When some eigenvalues of $\Abf$ are equal, the corresponding Schur polynomials~\eqref{eq:Schur} are obtained by using l'Hospital's rule.}. Schur polynomials form a basis in the space of homogeneous symmetric polynomials in $n$ variables of degree $k$ for all $k\leq n$. In particular, we have~\cite[Eq.~(17)]{1964James}
\begin{equation}\label{eq:TF}
(\Tr \Abf)^{k}=\sum_{\kappa}C_{\kappa}(\Abf).
\end{equation}

We will need the following identity\footnote{$\etr(\cdot)=e^{\Tr\left(\cdot\right)}$ denotes exponential of trace.}~\cite[Eq.~(6.2.3)]{1997Mathai}
\begin{multline}
\int_{\Xbf}\etr\left(-\Xbf \Zbf\right)\left(\det\left(\Xbf\right)\right)^{n-p} \\
\times ~_{r}\widetilde{F}_{s}\left(a_{1},\dots,a_{r};b_{1},\dots,b_{s};\Xbf\right) d \Xbf = \\
_{r+1}\widetilde{F}_{s}\left(a_{1},\dots,a_{r},n;b_{1},\dots,b_{s};\Zbf^{-1}\right)\left(\det\left(\Zbf\right)\right)^{-n}\widetilde{\Gamma}_{p}(n),
\label{eq:identity}
\end{multline}
where $\Xbf$, $\Zbf$ are $p\times p$ Hermitian matrices and $\widetilde{\Gamma}_{p}(n)$ is defined in \eqref{eq:MultiGammaC}.

We will prove the following lemma.
\begin{Lem} \label{Lem:Lu}
For any $p\times n$ complex matrix $\Sbf$, we have
\begin{equation}
\int_{\Ybf \in \stief{C}{n}{p}}\etr\left(2\mathfrak{R}\left(\Sbf \Ybf\right)\right)d\mu(\Ybf)= ~_{0}\widetilde{F}_{1}\left(n;\Sbf \Sbf^H\right).
\end{equation}
where $d\mu(\Ybf)$ is the uniform measure.
\end{Lem}

\par\noindent\textit{Proof:} First, note that it is equivalent to 
show that
\begin{multline}\label{eq:LR}
\left(\det\Sbf\Sbf^H\right)^{n-p}\int_{\Ybf}\etr\left(\Sbf \Ybf+\Ybf^H\Sbf^H\right)d\mu(\Ybf)= \\ \left(\det\Sbf\Sbf^H\right)^{n-p}~_{0}\widetilde{F}_{1}\left(n;\Sbf \Sbf^H\right).
\end{multline}
We will show that the matrix-variate Laplace transforms of both sides of the above equation are the same. The Laplace transform of the left-hand-side (LHS) is
\begin{eqnarray}\label{eq:dS}
T_{\text{LHS}}(\Zbf) \!\!&\!\!=\!\!&\!\!\int_{\Sbf\Sbf^H}\etr\left(-\Sbf\Sbf^H\Zbf\right)\left(\det\Sbf\Sbf^H\right)^{n-p}\nonumber\\
  && \times \int_{\Ybf}\etr\left(\Sbf \Ybf+\Ybf^H\Sbf^H\right)d\mu(\Ybf) d\left(\Sbf\Sbf^H\right)\nonumber\\
\!\!&\!\!=\!\!&\!\! \frac{\widetilde{\Gamma}_{p}(n)}{\pi^{np}}\int_{\Ybf}\int_{\Sbf}\etr\left(-\Sbf\Sbf^H\Zbf    \right.  \nonumber\\
 &  & \quad\quad\quad\quad \left.  +\Sbf \Ybf+\Ybf^H\Sbf^H\right)d\Sbf d\mu(\Ybf).
\end{eqnarray}
The 
equality above 
is established by utilizing the decomposition~\cite[Th.~4.5]{1997Mathai} $
\Sbf=\Lbf\Ubf$ 
where $\Lbf$ is a $p\times p$ lower triangular matrix with positive
diagonal elements and $\Ubf^H \in \stief{C}{n}{p}$, 
which leads to the fact that~\cite[Coroll.~4.5.3]{1997Mathai}
\begin{equation}
d\Sbf=2^{-p}\left(\det\Sbf\Sbf^H\right)^{n-p}d\left(\Sbf\Sbf^H\right)d\nu_{\Ubf},
\end{equation}
where here~\cite[Coroll.~4.5.2]{1997Mathai}  the measure on the Stiefel manifold is normalized so that $\int_{\stief{C}{n}{p}} d\nu_{\Ubf}=\frac{2^{p}\pi^{np}}{\widetilde{\Gamma}_{p}(n)}$. 
Applying the transform $\Sbf=\Zbf^{-1/2}\Tbf$ with
$
d\Sbf=\left(\det\Zbf\right)^{-n}d\Tbf
$
in~(\ref{eq:dS}), we have
\begin{eqnarray}
T_{\text{LHS}}(\Zbf)\!\!&\!\!=\!\!&\!\! \frac{\widetilde{\Gamma}_{p}(n)\left(\det\Zbf\right)^{-n}}{\pi^{np}}\int_{\Ybf}\int_{\Tbf}\etr\left(-\mathbf{TT}^H
 \right. \nonumber\\
 & &  \left. +\Zbf^{-1/2}\mathbf{TY}+\Ybf^H\Tbf^H(\Zbf^{-1/2})^H\right)d\Tbf d\mu(\Ybf) \nonumber\\
\!\!&\!\!=\!\!&\!\! \frac{\widetilde{\Gamma}_{p}(n)\left(\det\Zbf\right)^{-n}\etr\left(\Zbf^{-1}\right)}{\pi^{np}}\times\nonumber\\
 & & \quad \int_{\Ybf}\int_{\Tbf}\etr\left(-\left(\Tbf-\Ybf^H(\Zbf^{-1/2})^H\right)\right. \nonumber\\
  & & \quad \quad\quad \quad  \left.\left(\Tbf- \Ybf^H(\Zbf^{-1/2})^H\right)^H\right)d\Tbf d\mu(\Ybf) \nonumber\\
\!\!&\!\!=\!\!&\!\!  \widetilde{\Gamma}_{p}(n)\left(\det\Zbf\right)^{-n}\etr\left(\Zbf^{-1}\right),
\end{eqnarray}
where the last step is established by the fact that
$
p(\Xbf)=\frac{1}{\pi^{np}}\etr\left(-\left(\Xbf-\mathbf{M}\right)\left(\Xbf-\mathbf{M}\right)^H\right)
$
is a matrix-variate Gaussian density function.

The Laplace transform of the right-hand side (RHS) of~(\ref{eq:LR}) is
\begin{eqnarray}
T_{\text{RHS}}(\Zbf)\!\!&\!\!=\!\!&\!\!\int_{\Sbf \Sbf^H}\etr\left(-\Sbf\Sbf^H\Zbf\right)\left(\det\left(\Sbf\Sbf^H\right)\right)^{n-p}\nonumber\\
& & \quad \quad  \quad \quad \times ~_{0}\widetilde{F}_{1}\left(n;\Sbf \Sbf^H\right)d\left(\Sbf\Sbf^H\right) \nonumber\\
\!\!&\!\!=\!\!&\!\!\widetilde{\Gamma}_{p}(n)\left(\det\left(\Zbf\right)\right)^{-n}~_{1}\widetilde{F}_{1}\left(n;n;\Zbf^{-1}\right)\label{eq:1F1}\nonumber\\
\!\!&\!\!=\!\!&\!\!\widetilde{\Gamma}_{p}(n)\left(\det\left(\Zbf\right)\right)^{-n}~_{0}\widetilde{F}_{0}\left(\Zbf^{-1}\right)\nonumber\\
\!\!&\!\!=\!\!&\!\!\widetilde{\Gamma}_{p}(n)\left(\det\left(\Zbf\right)\right)^{-n}\etr\left(\Zbf^{-1}\right)\nonumber\\
\!\!&\!\!=\!\!&\!\! T_{\text{LHS}}(\Zbf),
\end{eqnarray}
where the second equality is obtained by the identity~(\ref{eq:identity}). By the uniqueness of Laplace transforms, we complete the proof of the lemma. 

\subsection{Stiefel Manifold} 
We now prove Lemma~\ref{lem:ChDistGaussian} in the case of the Stiefel manifold. 
The expansion of the chordal distance in terms of an inner product in the ambient space gives  
\begin{equation}
d_c^2(\Ibf_{n,p},\Ybf) = \| \Ibf_{n,p}-\Ybf \|^2_{F} = 2p - 2 \mathfrak{R} \Tr(\Ibf_{n,p}^H \Ybf).
\label{eqAp:distStief}
\end{equation}    
Accordingly, define the linear statistic  
\begin{equation}
Y_n =  \sqrt{\frac{2n}{p}} \mathfrak{R} \Tr(\Ibf_{n,p}^H \Ybf), 
\label{eq:LSStief}
\end{equation}
so that the distance is \mbox{$d_c^2(\Ibf_{n,p},\Ybf) = 2p\left(1 -\frac{1}{\sqrt{2pn}} Y_{n}\right)$}.

This type of linear statistic  converges in distribution to a standard Gaussian random variable as $n$ approaches infinity~\cite{Richards}.  
To see this, consider the moment-generating function of $Y_n$ which can be  represented as a hypergeometric function of matrix argument using Lemma~\ref{Lem:Lu}  with $ \Sbf = \sqrt{\frac{n}{2p}} \nu  \Ibf_{n,p}^H$,
\begin{eqnarray}
\expect{e^{\nu Y_n}} &=& ~_{0}\widetilde{F}_{1}\left(n; \frac{n \nu^{2}}{2p}     \Ibf_{n,p}^H \Ibf_{n,p}\right) \label{Eq:(106)}  \nonumber \\
&=& ~_{0}\widetilde{F}_{1}\left(n;\frac{n \nu^{2}}{2p}\Ibf_{p}\right) . 
\end{eqnarray}
Note that it corresponds to  the partition function of the von-Mises Fisher distribution~\cite{chikuse2012statistics} with parameter matrix $\sqrt{\frac{2 n }{p}} \nu \Ibf_{n,p}$.

By the definition of the hypergeometric function~\eqref{eq:MHF}, we can further write 
\begin{eqnarray}
\expect{e^{\nu Y_n}} &=&\sum_{k=0}^{\infty}\sum_{\kappa}\frac{1}{(n)_{\kappa}}\frac{C_{\kappa}\left(\frac{n \nu^{2}}{2p} \Ibf_{p}\right)}{k!}  \nonumber\\
&=&\sum_{k=0}^{\infty}\frac{\left(\frac{n \nu^{2}}{2p} \right)^{k}}{k!}\sum_{\kappa}\frac{C_{\kappa}(\Ibf_{p})}{(n)_{\kappa}}\label{eq:DnHS},
\end{eqnarray}
where the last equality is established by~(\ref{eq:Schur}). Since the leading order term in $(n-j+1)_{\kappa_{j}}$ equals $n^{\kappa_{j}}$, by~(\ref{eq:MHC}), for large $n$ we have
\begin{equation}
(n)_{\kappa}=\prod_{j=1}^{n}(n-j+1)_{\kappa_{j}}\sim n^{\kappa_{1}+\dots+\kappa_{n}}=n^{k}.
\end{equation}
Using~(\ref{eq:TF}), the sum in~(\ref{eq:DnHS}) for large $n$ becomes
\begin{equation}\label{eq:FP}
\sum_{\kappa}\frac{C_{\kappa}(\Ibf_{p})}{(n)_{\kappa}}\overset{n\to\infty}{=}\frac{1}{n^{k}}\sum_{\kappa}C_{\kappa}(\Ibf_{p})=\frac{1}{n^{k}}\Tr^{k}(\Ibf_{p})=\frac{p^{k}}{n^{k}},
\end{equation}
and we arrive at the result,
\begin{equation}
\lim_{n\to\infty}\expect{e^{\nu Y_n}}=\sum_{k=0}^{\infty}\frac{\left(\frac{\nu^{2}}{2}\right)^{k}}{k!}=e^{\frac{\nu^{2}}{2}},
\end{equation}
which is the moment-generating function of a zero mean and unit variance Gaussian distribution. 

Furthermore, we can directly identify in~\eqref{eq:DnHS} the finite-size moments of $Y_n$ from the series expansion of the moment-generating function $\expect{e^{\nu Y_n}} = \sum_{l=0}^{\infty}   \frac{\nu^l}{l!}\expect{Y_n^l }$. For any $n$, the mean is $\expect{Y_n}=0$  since there are no odd powers of $\nu$ in~\eqref{eq:DnHS}. The variance follows from the $(k=1)$-term in~\eqref{eq:DnHS} for which there is only one partition $\kappa=\{1,0,\ldots,0\}$  such that $(n)_{\kappa} = n$,  $\chi_{\kappa}(1)=1$, $C_{\kappa}(\Ibf_{p})=\chi_{\kappa}(\Ibf_{p})=p$, and thus $\var{Y_n} = 1$ also for any finite $n$.
 
As a byproduct, this convergence can be written in terms of the chordal distance  as
\begin{equation}
\sqrt{\frac{n}{2p}} \left(d_c^2(\Ibf_{n,p},\Ybf)-2p\right) \xrightarrow{d} \Ncal\left(0,1 \right)  \text{ as } n\to \infty ,
\end{equation} 
and where the finite-size regime mean and variance are exactly $\expect{d_c^2(\Ibf_{n,p},\Ybf)}= 2p$ and $\var{d_c^2(\Ibf_{n,p},\Ybf)}= \frac{2p}{n}$.

\subsection{Grassmann Manifold}

The case of the Grassmann manifold can be deduced as a by product of
the volume computation in~\cite{PWTC15} by setting $q=p$ (i.e. the
dimension of the center of the ball and the elements in the ball have
the same dimension). We provide below alternative lines of derivation
from the hypergeometric function interpretation consistent with the
Stiefel case discussed above.

If $\Ybf$ is uniformly distributed on the Stiefel manifold, then $[\Ybf]$ is uniformly distributed on the Grassmann manifold. 
A point $[\Ybf]$ on the Grassmann manifold can be uniquely defined by its projection matrix $\Pi_{\Ybf}= \Ybf \Ybf^H$. 
Looking at the uniform measure as a probability measure, the mapping  $\Ybf \to \Pi_{\Ybf}$ maps the uniform distribution on  $\stief{C}{n}{p} $ to the uniform distribution on $\grass{C}{n}{p}$. 
Namely, if $\Ybf$ has the same distribution as  $\Ubf \Ybf$ for all $\Ubf \in \Ucal_n$, then $\Pi_{\Ubf\Ybf}$ has the same distribution as  $\Ubf \Pi_{ \Ybf}\Ubf^H$ for all $\Ubf \in \Ucal_n$. 
Accordingly, consider 
\begin{eqnarray}
d_c^2([\Ibf_{n,p}],[\Ybf]) &=& \frac{1}{2}\| \Ebf-\Pi_{\Ybf}\|^2_{F} \nonumber\\
&=& p - \Tr( \Ybf^H \Ebf \Ybf)
\label{eqAp:distGrass}
\end{eqnarray}
with $\Ebf=\Ibf_{n,p}\Ibf_{n,p}^H$ and such that $\Ybf$ is uniformly distributed on the Stiefel manifold $\stief{C}{n}{p}$.	 

The chordal distance can thus be expressed 
as a function of a linear statistic  
$Z_{n,p} =  \Tr( \Ybf^H \Ebf \Ybf)$.  
This type of linear statistic  is asymptotically Gaussian for $n,p \to \infty $ and $n-2p$ constant as shown in~\cite{PWTC15}\footnote{One has $Z_{n,p}= \sum_{i=1}^p \cos^2 \theta_i$ from the definition of the chordal distance~\eqref{GrassChorDistCos}. Computing the moment-generating function, one gets $\expect{e^{\nu Z_{n,p} }} = e^{p\nu} D_p(i \nu)$ where $D_p(\nu)$ is defined in~\cite[Eq.~(22)]{PWTC15}. } 
through its interpretation  as the sum of squared principal cosines distributed according to the Jacobi ensemble~\cite[Sec. 2.1.2.]{johnstone2008multivariate} and using~\cite[Th. 2]{Johansson1997}. 

The moment-generating function of $Z_{n,p}$ corresponds to the partition function of  the Bingham distribution~\cite{chikuse2012statistics} with parameter matrix $\nu \Ebf$, which  can be expressed as a confluent hypergeometric function of complex matrix argument. This can be verified by matrix-variate Laplace transforms, as in the proof of Lemma~\ref{Lem:Lu}. 
It follows that the moments of $Z_{n,p}$ can also be computed by identification from the definition of the  hypergeometric function~\eqref{eq:MHF}:  
\begin{eqnarray}
\expect{e^{\nu Z_{n,p} }} &=& \sum_{k=0}^{\infty}   \frac{\nu^k}{k!}\expect{Z_{n,p}^k } \nonumber\\
 &=& ~_{1}\widetilde{F}_{1}\left(p,n;\nu \Ebf \right) \nonumber\\
&=& \sum_{k=0}^{\infty} \frac{\nu^k}{k!} \sum_{\kappa} \frac{(p)_{\kappa}}{(n)_{\kappa}} C_{\kappa}(\Ebf).
\end{eqnarray}

The  mean of $Z_{n,p}$ is given for $k=1$ for which there is only one partition such that $\chi_{\kappa}(\Ebf) = p$, and 
$
\expect{Z_{n,p}} = \frac{p^2}{n}. 
$ 
For the second moment with $k=2$, the possible partitions are $\kappa=\{2,0,\ldots,0\}$ and $\kappa=\{1,1,0,\ldots,0\}$. The corresponding  Schur polynomials  for each partition are  $\chi_{\kappa}(\Ebf)=\frac{p}{2}(p+1)$ and $\chi_{\kappa}(\Ebf)=\frac{p}{2}(p-1)$, respectively; while for both partitions one has $\frac{(p)_{\kappa}}{(n)_{\kappa}}=\frac{p(p+1)}{n(n+1)}$ and $\chi_{\kappa}(1)=1$. One obtains after some manipulations 
$
\expect{Z_{n,p}^2} = \frac{p^2(np^2-2p+n)}{n(n^2 -1)}
$ and thus the variance $\var{Z_{n,p}}=  \frac{p^2(n-p)^2 }{n^2 (n^2-1)} $.
Therefore, we have 
$$\frac{n \sqrt{n^2-1}} {p(n-p) } \left(Z_{n,p} -\frac{p^2}{n}\right) \xrightarrow{d} \Ncal\left(0,1 \right). $$  
Equivalently, the squared chordal distance for the Grassmann manifold converges in distribution to a Gaussian random variable as 
$$\frac{n \sqrt{n^2-1} }{p(n-p)}   \left( d_c^2([\Ibf_{n,p}],[\Ybf]) - \frac{p(n-p)}{n}\right) \xrightarrow{d} \Ncal\left(0, 1\right)$$ with $n,p \to \infty $ and $n-2p$ constant, and its finite-regime mean and variance are exactly $\expect{d_c^2([\Ibf_{n,p}],[\Ybf])} = \frac{p(n-p)}{n}$ and  $\var{d_c^2([\Ibf_{n,p}],[\Ybf])}  = \frac{p^2(n-p)^2}{ n^4-n^2}$.

\section{Proof of the Volume Comparison in Theorem~\ref{thm:cap_asympt}}
\label{proof:VCTheorem}

Now with Lemma~\ref{lem:ChDistGaussian}, we are in position to prove Theorem~\ref{thm:cap_asympt}. The result follows by identification of the asymptotic forms between the manifold and the embedding sphere.  

\subsection{Stiefel Manifold}
By choosing without loss of generality $\Ibf_{n,p}$ as the center of the ball and 
noting that  
$$
0 \leq d_c(\Ibf_{n,p},\Ybf) \leq r ~~\Leftrightarrow~~ 
 \sqrt{2 n p } \geq  Y_n \geq  \sqrt{\frac{n}{2p}}(2p-r^2)
$$ 
where $Y_n = \sqrt{\frac{n}{2p}}\left(2p-d_c^2(\Ibf_{n,p},\Ybf)\right) $; identification of the normalized volume of the ball as a probability measure gives 
\begin{eqnarray} 
\!\!\!\!\!\! \mu(B(r)) \!\!&\!\!=\!\!&\!\! {\rm Pr}\{ \Ybf \in \stief{C}{n}{p} \, | \, 0 \leq d_c(\Ibf_{n,p},\Ybf)\leq r\} \nonumber \\				 
					\!\!&\!\!=\!\!&\!\! F_{Y_n}\left(\sqrt{2 n p } \right) - F_{Y_n} \!\!\left(\!\!\sqrt{2np}\left(1-\frac{r^2}{2p}\right)\!\!\right)	
									\label{eq:ballStiefFdc}
\end{eqnarray} 
where $F_{Y_n}(z) $ is the cumulative distribution of the random variable $Y_n$ with $\Ybf$ being uniformly distributed on $\stief{C}{n}{p}$. Note that $F_{Y_n}(\sqrt{2 n p } ) = 1 $ by definition of the chordal distance. 

As shown via Lemma~\ref{lem:ChDistGaussian}, $Y_n$  converges in distribution to a standard Gaussian random variable, and thus  with  $n \to \infty$ and $z= \sqrt{2np}\left(1-\frac{r^2}{2p}\right) $ fixed, one has 
\begin{eqnarray} 
\lim_{n\to \infty }\mu(B(r))\!\!&\!\!=\!\!&\!\!  1- \lim_{n\to \infty }F_{Y_n}\left(z\right) \nonumber \\
                    \!\!&\!\!=\!\!&\!\!  \frac{1}{2}- \frac{1}{2} \erf\left(\sqrt{np}\left(1-\frac{r^2}{2p}\right) \right) . \label{eq:ballStiefFdcInf}
\end{eqnarray} 
The final asymptotic expression is obtained by using  the finite-size correction  
$F_{Y_n}(\sqrt{2 n p } )\approx  \frac{1}{2} \left(1+ \erf\left(\sqrt{ n p }  \right) \right) \to 1$ in~\eqref{eq:ballStiefFdc} in order to provide an approximation that simulateneously satisfies  $\mu(B(0))=0$ and the asymptotic limit~\eqref{eq:ballStiefFdcInf}. 
This leads to~\eqref{eq:vol_asymp_stief1}. 

Finally, one verifies  that the asymptotic volume of a ball in~\eqref{eq:vol_asymp_stief1} is exactly the asymptotic volume of a cap in the embedding sphere   $\stief{C}{n}{p} \hookrightarrow \Scal^{2np-1}(\sqrt{p})$ by identification with $D = 2np$ and $R=\sqrt{p}$ in~\eqref{eq:largecap}.

\subsection{Grassmann Manifold}
Again from the results of~\cite{PWTC15}, one may directly obtain~\eqref{eq:vol_asymp_grass}  
by setting $q=p$. 
Alternatively, the volume of a ball of radius $r$ centered around  $[\Ibf_{n,p}]$ can be expressed as 
\begin{eqnarray} 
\!\!\!\!\!\! \mu(B(r)) \!\!&\!\!=\!\!&\!\! {\rm Pr}\{ [\Ybf]  \in \grass{C}{n}{p} \, | \, 0 \leq d_c([\Ibf_{n,p}],[\Ybf])\leq r\} \nonumber\\
					\!\!&\!\!=\!\!&\!\! {\rm Pr}\{ \Ybf  \in \stief{C}{n}{p} \, | \, 0 \leq d_c([\Ibf_{n,p}],[\Ybf])\leq r\} \nonumber\\ 
									\!\!&\!\!=\!\!&\!\! F_{\tilde{Z}_{n,p}}\left(\sqrt{n^2-1} \right ) \nonumber\\
									\!\!& \!\!& \quad \!\!  -F_{\tilde{Z}_{n,p}}\!\!\left(\!\!\sqrt{n^2-1} \left(1-\frac{n \ r^2}{p(n-p)}\right)\!\!\right),  \label{eq:ballGrassFdc}
\end{eqnarray} 
where $F_{\tilde{Z}_{n,p}}(z) $ is the cumulative distribution of the random variable $\tilde{Z}_{n,p}=\frac{n \sqrt{n^2-1} }{p(n-p)}   \left( \frac{p(n-p)}{n} - d_c^2([\Ibf_{n,p}],[\Ybf])\right) $ such that $\Ybf$ is uniformly distributed on the Stiefel manifold $\stief{C}{n}{p}$.	 

As shown in Lemma~\ref{lem:ChDistGaussian}, $\tilde{Z}_{n,p}$ converges in distribution to a standard Gaussian random variable as $n,p \to \infty$ with $(n-2p)$ fixed.  Thus  with $z= \sqrt{n^2-1} \left(1-\frac{n \ r^2}{p(n-p)}\right) $  also fixed, one has 
\begin{eqnarray} 
\lim_{n,p\to \infty }\mu(B(r)) \!\!&\!\!=\!\!&\!\!1- \lim_{n,p\to \infty }F_{\tilde{Z}_{n,p}}\left(z)\right) \nonumber \\
                    \!\!&\!\!=\!\!&\!\! \frac{1}{2}- \frac{1}{2} \erf\left(\sqrt{\frac{n^2-1}{2}} \left(1-\frac{n \ r^2}{p(n-p)}\right) \right) . \nonumber\\
										&& \label{eq:ballGrassFdcInf}
\end{eqnarray} 
Again, the final asymptotic expression~\eqref{eq:vol_asymp_grass} is obtained by using the finite-size correction  
$F_{\tilde{Z}_{n,p}}\left(\sqrt{n^2-1} \right ) \approx  \frac{1}{2} \left(1+ \erf\left(\sqrt{\frac{n^2-1}{2}} \right) \right) \to 1$ in~\eqref{eq:ballGrassFdc}    in order to provide a volume approximation that simultaneously satisfies $\mu(B(0))=0$ and the asymptotic condition~\eqref{eq:ballGrassFdcInf}. 

Finally, one verifies   also here that the asymptotic volume of a ball in~\eqref{eq:vol_asymp_grass} equals the asymptotic volume of a cap in the embedding sphere  $\grass{C}{n}{p} \hookrightarrow \Scal^{n^2-2}\left(\sqrt{\frac{p(n-p)}{2n}}\right)$ by identification with $D = n^2-1$ and $R=\sqrt{\frac{p(n-p)}{2n}}$ in~\eqref{eq:largecap}.

\section{Proof of Kissing Radius Bounds in  Proposition~\ref{prop:kissing_radius}}
\label{proof:kissing}
Consider a code $\Ccal= \{\Cbf_1,\ \ldots, \Cbf_N\}$ with pairwise distances among the codewords   $\{ \delta_{i,j}\}_{i \neq j}$ such that $\delta = \min \{ \delta_{i,j}\} $. Between each codeword there is a different  mid-distance $\{ \varrho_{i,j}\}_{i \neq j}$ and the kissing radius $\varrho = \min{\varrho_{i,j}}$. The detailed derivations below provide a lower and upper bound on a mid-distance  $\rhol(\delta_{i,j})   \leq \varrho_{i,j} \leq   \rhou(\delta_{i,j}) $ as function of the distance $\delta_{i,j}$.  It can be verified that the obtained bounds $\rhol$ and $\rhou$ are increasing functions. It then follows that $\min \{\rhol(\delta_{i,j}) \}= \rhol(\min\{\delta_{i,j}\})  $ and  $\min\{ \rhou(\delta_{i,j})\} = \rhou(\min\{\delta_{i,j}\})  $, and one has $\rhol(\delta) \leq \varrho \leq \rhou(\delta)$.

\subsection{Grassmann Manifold}
The principal angles $\thetabf = (\theta_1,\ldots,\theta_p)$ between two codewords $\Cbf_i,\ \Cbf_j$ separated by $\delta$ satisfies $\sum_{i=1}^{p} \sin^2(\theta_i)=\delta^2$. 
Without loss of generality, the code may be rotated so that the Stiefel representatives of these codewords are of the form $\Cbf_i= \Ibf_{n,p} $ and $\Cbf_j=[(\mbox{diag} ( \cos \thetabf) \, \mbox{diag} ( \sin \thetabf))^T ] $~\cite{conway-1996-5}.

The chordal distance is measured along the geodesic. The principal angles between the midpoint $\Mbf_{i,j}$ (on the geodesic joining $\Cbf_i$ and $\Cbf_j$) and a codeword $\Cbf_i$ or $\Cbf_j$  are $(\frac{\theta_1}{2},\ldots,\frac{\theta_p}{2})$~\cite{Edelman}. It follows that the squared chordal distance between the midpoint on the geodesic and an extremity of the geodesic is 
\begin{equation}\varrho^2= \left\| \sin{\frac{\thetabf}{2}} \right\|^2_2=\sum_{i=1}^{p} \sin^2 \frac{\theta_i}{2} .
\end{equation}

Finding lower and upper bounds reduces to solving the following optimization problems:  
\begin{equation}
\label{eq:constraintP}
  \begin{array}{rcl}
	 \underset{\thetabf \in [0,\ \frac{\pi}{2}]^p}{\text{minimize/maximize}}
	&~& \| \sin{\frac{\thetabf}{2} } \|^2_2 \\
	\text{subject to}
	&~&  \| \sin{\thetabf} \|^2_2=\delta^2.
	\end{array}
\end{equation}

First, to find the minimum,  consider the corresponding Lagrange function 
\begin{equation}
\Lambda(\theta_1,\ldots,\theta_p,\lambda) = \| \sin{\frac{\thetabf}{2}}\|^2_2  + \lambda \Big(  \| \sin{\thetabf} \|^2_2-\delta^2   \Big) .
\end{equation}
Solving:
\begin{eqnarray}
\!\!\!\!\!\!\!\!\!\!\!\! \frac{\partial \Lambda}{\partial \theta_i}  \!\!&\!\!=\!\!&\!\! \sin \theta_i(1/2+2 \lambda \cos \theta_i) = 0  \quad  \text{for  } i=1 \ldots p\\
\!\!\!\!\!\!\!\!\!\!\!\! \frac{\partial \Lambda}{\partial \lambda}  \!\!&\!\!=\!\!&\!\!  \sum_{i=1}^{p} \sin^2 \theta_i-\delta^2 =0
\end{eqnarray}
yields a set of stationary points where at least $x$ angles are nonzero such that $x \geq \lceil \delta^2 \rceil$ and equal to $\theta^*=\arcsin\frac{\delta}{\sqrt{x}}$.
It is then easy to verify that the objective funtion $f(x)=\sum_{i=1}^{p} \sin^2 \frac{\theta_i}{2}= x/2(1-\sqrt{1-\delta^2/x})$ is a strictly decreasing function on $[\lceil \delta^2 \rceil, \,p]$
and thus is minimized for $x=p$. The result follows.

Maximization in \eqref{eq:constraintP} is obtained when a minimum number of angles is maximized, i.e.,  with $(\theta_1^\star,\ldots,\theta_p^\star) \in [0,\ \frac{\pi}{2}]^p$ such that 
$\theta_1^\star=\cdots=\theta_{\lfloor \delta^2 \rfloor}^\star=\frac{\pi}{2}$, 
$\theta_{\lceil \delta^2 \rceil}^\star=\arcsin(\sqrt{\delta^2-\lfloor \delta^2 \rfloor} )$ and $\theta_{\lceil \delta^2 \rceil+1}^\star=\cdots=\theta_p^\star=0$. 
This can be verified by contradiction as in~\cite{Han}: 
Defining $s_i= \sin^2\theta_i$ and  $t(s_i)= (1-\sqrt{1-s_i})/2$, consider the equivalent problem of maximizing $\sum t(s_i)$ such that $\sum s_i= \delta^2$ and without loss of generality $ 1 \geq s_1 \geq s_2 \cdots \geq s_p \geq 0$.
By contradiction, assume that  $\sum t(si)$ is maximum at $\abf$ with $a_i>0 \, \forall i$. 
It is possible to find a $\bbf$ with $b_i \geq a_i$ for $1\leq i \leq p-1$ and  $b_p=0$. 
Since $t'(\cdot)$ is strictly increasing, it follows from the mean value theorem that there exist $c\in (a_{p-1},\,b_{p-1})$ and $d\in (0,\,a_p)$ such that
\begin{eqnarray}
\sum t(b_i)-t(a_i) &\geq& t'(c) \sum_{i=1}^{p-1} (b_i-a_i)+ t'(d)(a_p-b_p) \nonumber\\
 &=& (t'(d)+t'(c))a_p >0,
\end{eqnarray}
where the last equality is due to the constraint  $\sum b_i= \sum a_i= \delta^2$, which is in contradiction with the fact that $\sum t(a_i)$ is a maximum. Repeating the procedure from $s_p$ to $s_{\lceil \delta^2 \rceil}$ leads to the results.
Lastly, the maximum is 
\begin{eqnarray}
\sum_{i=1}^{p} \sin^2 \frac{\theta_i^\star}{2} &=& \frac{ \lfloor \delta^2 \rfloor}{2}+ \frac{ 1- \sqrt{1- ( \delta^2-\lfloor \delta^2 \rfloor) }}{2} \nonumber\\
			&=& \frac{1}{2} \left(\lceil \delta^2 \rceil - \sqrt{ \lceil \delta^2 \rceil-\delta^2} \right).
\end{eqnarray}

\subsection{Unitary Group}
A simple adaptation of the proof for the unitary group can be done, see also~\cite{Han} where similar optimization problems are considered. 
Consider a unitary code $\Ccal$ with minimum distance $\delta$. 
The angles $(\theta_1,\ldots,\theta_n)$ between two codewords $\Cbf_i,\ \Cbf_j$ separated by $\delta$ now satisfies $\sum_{i=1}^{n} \sin^2(\frac{\theta_i}{2})=\frac{\delta^2}{4}$. 
Again, the chordal distance is measured along the geodesic and the principal angles between the midpoint and a codeword  are $(\frac{\theta_1}{2},\ldots,\frac{\theta_n}{2})$. It follows that the squared chordal distance between the midpoint on the geodesic and an extremity of the geodesic is 
\begin{equation} 
\varrho^2= 4\sum_{i=1}^{n} \sin^2 \frac{\theta_i}{4} . \end{equation}

Finding lower and upper bounds reduces to solving the following optimization problems:  
\begin{equation}
\label{eq:constraintPUnitary}
  \begin{array}{rcl}
	 \underset{\thetabf \in [-\pi,\ \pi]^n}{\text{minimize/maximize}}
	&~& 4 \| \sin{\frac{\thetabf}{4}} \|^2_2 \\
	 \text{subject to}
	&~&  \| \sin{\frac{\thetabf}{2}} \|^2_2=\frac{\delta^2}{4}.
	\end{array}
\end{equation}

By using the change of variables $\phibf  = (\thetabf +\pi)/2 $, $\gamma = \frac{\delta}{2} $ and $\rho = \frac{\varrho}{2} $, we recover the optimization problem~\eqref{eq:constraintP}. The result follows.

\subsection{Stiefel Manifold}
For  a generic Stiefel manifold, the notion of principal angles does not exist. The obtained lower bound for the unitary group is actually matching the kissing radius bound from the spherical embedding. This result thus directly extends to all Stiefel manifolds. However, we were not able to generalize the upper bound. We provide some discussion below.

Without loss of generality, we can assume that the first point is $\Ibf_{n,p}$ and the second point is $\Ybf$. If $\Ybf$ is block unitary as $\Ybf = [\Ubf\; \bm{0}]^T$, where $\Ubf \in \Ucal_p$ then the geodesic between this two points stays in $\Ucal_p$ embedded in $\stief{C}{n}{p}$, and the upper bound from the unitary group would apply. 

For a generic $\Ybf$, the distance only depends on the diagonal element of $\Ybf = \{Y_{ij} \}$:
\begin{eqnarray}
d_c^2(\Ibf_{n,p}, \Ybf) \!\!&\!\!=\!\!&\!\! 2(p- \mathfrak{R}(\mathrm{Tr}(\Ibf_{n,p}^H\Ybf)) = 2\left(p- \sum_{i=1}^p \mathfrak{R}(Y_{ii})\right) \nonumber\\
								\!\!&\!\!=\!\!&\!\! 2\left(p- \sum_{i=1}^p \cos \theta_i \right)=4\sum_{i=1}^p \sin^2 \frac{\theta_i}{2} \nonumber\\
									\!\!&\!\!=\!\!&\!\!  4p \sin^2 \frac{\theta}{2},
\end{eqnarray}
where without of loss of generality, we have written $\cos \theta_i = \mathfrak{R}( Y_{ii})$ and $\cos \theta = \frac{1}{p} \sum_{i=1}^p \cos \theta_i $.
The angles $\{\theta_i\}$ then correspond the canonical embbeding of the Stiefel manifold in $(\Scal^{2n-1})^p$, the angle $\theta$ to the embedding in $\Scal^{2np-1}(\sqrt{p})$. 

The geodesic is not along these angles, and the midpoint, say $\Mbf_{\Ybf}$, on the geodesic is not at $\{\frac{\theta_i}{2}\}$ nor at $\frac{\theta}{2}$. However since we have an isometry for these embeddings, 
\begin{eqnarray}
d_c^2(\Ibf_{n,p}, \Mbf_{\Ybf}) &\geq& 4\sum_{i=1}^p \sin^2 \frac{\theta_i}{4} \nonumber \\
								&\geq& 4p \sin^2 \frac{\theta}{4}.
\end{eqnarray}
From this, by maximizing the right hand-side of the first inequality, we can deduce that the maximum is greater or equal to the upper bound of the unitary group, i.e. 
\begin{equation}
\max_{\Ybf } d^2(\Ibf_{n,p}, \Mbf_{\Ybf}) \geq \rhou.
\end{equation}
A generalization would  imply that $\max_{\Ybf} d^2(\Ibf_{n,p}, \Mbf_{\Ybf}) = \rhou$.

\subsection{Proof of Corollary~\ref{Corr:ImprovedKRGrass}}
\label{Proof:ImprovedKRGrass}
For the Grassmann manifold with $R^2= \frac{p(n-p)}{2n}$, it can be easily verified that $p \leq 4R^2$ with equality if and only if $p=n/2$. Then, since $x/2(1-\sqrt{1-\delta^2/x})$ is a strictly decreasing function, it follows that $\varrho_s \leq \rhol$ with equality if and only if $p=n/2$.

\addcontentsline{toc}{chapter}{Bibliography}
\bibliographystyle{IEEEtran}
\bibliography{bounds}

\end{document}